\documentclass[aps,reprint,superscriptaddress]{revtex4-2}
\usepackage{siunitx}
\usepackage{graphicx}
\usepackage{placeins}
\usepackage{textcomp}
\usepackage{amssymb}
\usepackage{makecell}
\usepackage{physics}
\usepackage{textgreek}
\usepackage{natbib}
\usepackage{color}

\graphicspath{{Figures/}}
\begin{document}
\title{Magnetic properties of an individual \textit{Magnetospirillum gryphiswaldense} cell}
\author{Mathias M. Claus}
\affiliation{Swiss Nanoscience Institute, University of Basel, Klingelbergstrasse 82, 4056 Basel, Switzerland}
\affiliation{Department of Physics, University of Basel, Klingelbergstrasse 82, 4056 Basel, Switzerland}
\author{Marcus Wyss}
\affiliation{Swiss Nanoscience Institute, University of Basel, Klingelbergstrasse 82, 4056 Basel, Switzerland}
\author{Dirk Schüler}
\affiliation{Lehrstuhl für Mikrobiologie, University of Bayreuth, Universitätsstraße 30, 95447 Bayreuth, Germany}
\author{Martino Poggio}
\affiliation{Swiss Nanoscience Institute, University of Basel, Klingelbergstrasse 82, 4056 Basel, Switzerland}
\affiliation{Department of Physics, University of Basel, Klingelbergstrasse 82, 4056 Basel, Switzerland}
\author{Boris Gross}
\affiliation{Department of Physics, University of Basel, Klingelbergstrasse 82, 4056 Basel, Switzerland}
\date{\today}
\begin{abstract}
Many bacteria share the fascinating ability to sense Earth’s magnetic field –- a process known as magnetotaxis. These bacteria synthesize magnetic nanoparticles, called magnetosomes, within their own cell body and arrange them to form a linear magnetic chain. The chain, which behaves like a compass needle, aligns the microorganisms with the geomagnetic field. Here, we measure the magnetic hysteresis of an individual bacterium of the species \textit{Magnetospirillum gryphiswaldense} via ultrasensitive torque magnetometry. These measurements, in combination with transmission electron microscopy and micromagnetic simulations, reveal the magnetic configurations of the magnetosomes, their progression as a function of applied field, as well as the total remanent magnetic moment and effective magnetic anisotropy of a chain within a single bacterium. Knowledge of these properties is crucial both for understanding the mechanisms behind magnetotaxis and for the design of systems exploiting magnetotactic bacteria in biomedical applications.
%
\end{abstract}
\maketitle
\section{Introduction}
%

Nature has invented an incredible mechanism to lead magnetotactic bacteria such as \textit{Magnetospirillum gryphiswaldense} to its optimum feeding grounds: a compass needle in its cell body in the form of a chain of self-assembled magnetite crystallites, each a few tens of nanometers in diameter~\cite{uebe_magnetosome_2016, faivre_magnetotactic_2008, lefevre_ecology_2013}. Even better than a compass, which only indicates the direction of the Earth's magnetic field, this chain of magnetosomes directly orients the bacterium along the field, as a result of the torque exerted. Using its flagella and a biological sensor for oxygen content~\cite{popp_polarity_2014}, the bacterium then propels itself along the magnetic field lines towards sources of nourishment in aquatic sediments. By restricting \textit{M. gryph.}'s motion along one dimension, the chain of magnetosomes increases the efficiency of this search~\cite{frankel_magneto-aerotaxis_1997,schuler_biomineralization_2002}. 

The discovery of magnetotactic bacteria has inspired a number of ideas for biomedical applications~\cite{alphandery_use_2013, serantes_multiplying_2014,felfoul_magneto-aerotactic_2016,ghosh_m13-templated_2012,gandia_unlocking_2019,kraupner_bacterial_2017, mickoleit_versatile_2020}, including using them as nano-robots with magnetic actuation~\cite{khalil_chapter_2017,mishra_selective_2016} or as sensing devices~\cite{jiang_bioinspired_2016}. The biomineralization process governing the formation of magnetosome chains is also of great interest for finding new routes of material generation~\cite{schuler_genetics_2008}. More exotic research directions include the use of magnetotactic bacteria in waste water treatment~\cite{zheng_construction_2022} or studies linking the presence of magnetite nanoparticles -- similar to those making up magnetosomes -- in meteorites from Mars to potential presence of ancient life on that planet~\cite{arato_crystal-size_2005, thomas-keprta_magnetofossils_2002,frankel_magnetite_2000}.

The magnetism of a chain of magnetosomes within a single \textit{M. gryph.} is complex: It consists of several tens of individual nanomagnets, one after another in a string-like arrangement. Because of their small size, each magnetosome can be approximated as a single-domain magnet, whose magnetism is determined both by the cubic anisotropy of magnetite and the magnetosome's shape. The fact that each magnetosome takes an unknown orientation in space adds further complexity. Finally -- and crucially for the biological function of the chain -- dipolar interactions between nearby magnetosomes result in a net preferred orientation along the chain axis. The competition between these anisotropies defines the orientation of each magnetosome's magnetic moment.

Precise measurements of the magnetic properties of magnetotactic bacteria, specifically anisotropy and total magnetic moment, are the key to understanding the mechanisms behind magnetotaxis~\cite{frankel_magnetic_1984, kalmijn_biophysics_1981, zhu_angle_2014, klumpp_magnetotactic_2016, klumpp_swimming_2019, mohammadinejad_stokesian_2021, codutti_interplay_2022}. One approach is to monitor the motion of either ensembles or individual bacteria in the presence of magnetic fields~\cite{zahn_measurement_2017, reufer_switching_2014, le_sage_optical_2013, nadkarni_comparison_2013, bahaj_alternative_1996}. Reufer et al.~\cite{reufer_switching_2014} investigated the trajectories of \textit{M. gryph.} swimming in a magnetic field, determining an average magnetic moment of $2\cdot10^{-16}$~\si{\ampere\meter\squared} per bacterium. Zahn et al.~\cite{zahn_measurement_2017} extracted the moment of about 150 bacteria in liquid by tracking and modeling the rotation and translation of individual bacteria under the influence of magnetic tweezers. Typically, however, magnetic properties deduced from such measurements rely on assumptions about the motion of the bacteria in liquid. The magnetic moment of ensembles of magnetotactic bacteria has been measured directly using a superconducting quantum interference device (SQUID)~\cite{chemla_new_1999, prozorov_magnetic_2007}.
Individual bacteria have also been studied via magnetic imaging techniques, including X-ray magnetic circular dichroism combined with scanning transmission X-ray microscopy~\cite{lam_characterizing_2010, marcano_magnetic_2022, kalirai_examining_2012, kalirai_anomalous_2013, staniland_rapid_2007} and electron holography~\cite{dunin-borkowski_magnetic_1998}. In particular, Dunin-Borowski et al.\ used electron holography to visualize the magnetic stray fields of an individual \textit{Magnetospirillum magnetotacticum} in remanence~\cite{dunin-borkowski_magnetic_1998}.

The magnetic hysteresis of an individual bacterium, from which magnetic properties such as coercivity, anisotropy, and switching behavior can be determined, is challenging to measure due to the tiny magnetic moment of a single chain of magnetosomes~\cite{wasem_atomic_2015,orue_configuration_2018,marcano_magnetic_2022}.
Knowledge of these properties is especially important for applications involving magnetic actuation or sensing. 

Dynamic cantilever magnetometry (DCM)~\cite{rossel_active_1996, harris_integrated_1999, stipe_magnetic_2001, gross_dynamic_2016}, which is a particularly sensitive form of torque magnetometry, provides a method for measuring the magnetism of nanometer-scale magnetic systems. Gysin et al.\ first used DCM to measure the magnetic hysteresis of an ensemble of about 100 \textit{M. gryph.}, resulting in a magnetic moment of $5\cdot10^{-16}$~\si{\ampere\meter\squared}~\cite{gysin_magnetic_2011} per bacterium. A few years later, using a more sensitive cantilever, the same group managed to measure the hysteresis of an individual \textit{M. gryph.}~\cite{wasem_atomic_2015}, however, low signal-to-noise ratio precluded a detailed analysis of the measurements. 

By attaching a single \textit{M. gryph.\ }to an ultrasensitive SiN cantilever, we resolve its magnetic hysteresis via DCM. Furthermore, based on transmission electron microscopy (TEM) images, we create a micromagnetic model for the bacterium's chain of magnetosomes and compare it to DCM measurements. This comparison allows us to determine the likely progression of the magnetosomes' magnetic configurations as a function of applied field, as well as determining components of the remanent magnetization, effective anisotropy, and switching field of the bacterium. In measurements of ensembles, the properties of individual bacteria are obscured by inhomogeneity in the size, shape, and orientation of the chains of magnetosomes.
\section{Dynamic cantilever magnetometry}

In DCM, as shown in Fig.~\ref{fig:fig01}, the sample under investigation is attached to the free end of a cantilever, which is driven into self-oscillation at its fundamental mechanical resonance frequency $f$. The measurement consists of monitoring changes in this frequency $\Delta\!f = f - f_0$ as a function of a uniform applied magnetic field $\mathbf{H}$, where $f_0$ is the resonance frequency at $H = 0$. $\Delta\!f$ reveals the curvature of the magnetic system's free energy with respect to rotations about the cantilever oscillation axis~\cite{modic_resonant_2018,gross_magnetic_2021}:
$$\Delta f = \frac{f_0}{2k_0l_e^2} \left(\left.\frac{\partial^2E_m}{\partial\theta_c^2}\right|_{\theta_c=0}\right).$$
$E_m$ is the magnetic energy, $k_0$ the spring constant, $l_e$ the effective length, and $\theta_c$ the cantilever's oscillation angle. A detailed derivation of this relation can be found in the supplementary information of Ref.~\onlinecite{mehlin_stabilized_2015}. This relation does not allow a direct link between $\mathbf{m}$ and $\Delta f$, but $\Delta f$ can be intuitively understood in a thermodynamic framework~\cite{modic_resonant_2018} as an analogue to magnetic susceptibility: while susceptibility quantifies the magnetic response of a sample to changes in the magnitude of $\mathbf{H}$, $\Delta\!f$ quantifies the response to changes in its orientation. Measurements of $\Delta\!f$ are particularly useful for identifying magnetic phase transitions~\cite{mehlin_stabilized_2015,modic_resonant_2018,gross_stability_2020}, as well as providing information on the switching, saturation magnetization, coercivity, and the anisotropy of a magnetic system~\cite{philipp_magnetic_2021}.

The cantilevers used in this measurement are fabricated from SiN and are \SI{55.8}{\micro\meter} long, \SI{1.94}{\micro\meter} wide and \SI{50}{\nano\meter} thick. The resonance frequency $f_0$ of the fundamental mechanical mode used for magnetometry is \SI{16.9}{\kilo\hertz} with an effective spring constant of $\SI{44}{\micro\newton/\meter}$ and quality factor of a few thousand at room temperature. Once the sample has been attached to the end of the cantilever, the cantilever is mounted in a vibration-isolated vacuum chamber at $10^{-6}$~mbar. Using an external rotatable superconducting magnet, magnetic fields up to \SI{4.5}{\tesla} can be applied along any direction spanning \SI{225}{\degree} in the plane of cantilever oscillation ($xz$-plane), as specified by the angle $\theta_h$ shown in Fig.~\ref{fig:fig01}. The cantilever's flexural motion is read out using an optical fiber interferometer using \SI{100}{\nano\watt} of laser light at \SI{1550}{\nano\meter}~\cite{rugar_improved_1989}. A piezoelectric actuator mechanically drives the cantilever at $f_0$ with a constant oscillation amplitude of a few tens of nanometers using a feedback loop implemented by a field-programmable gate array. This process enables the fast and accurate extraction of $\Delta\!f$ from the cantilever deflection signal~\cite{SOM}.

\section{Sample preparation}
\begin{figure}[tb]
    \centering
\includegraphics[width=1.0\linewidth]{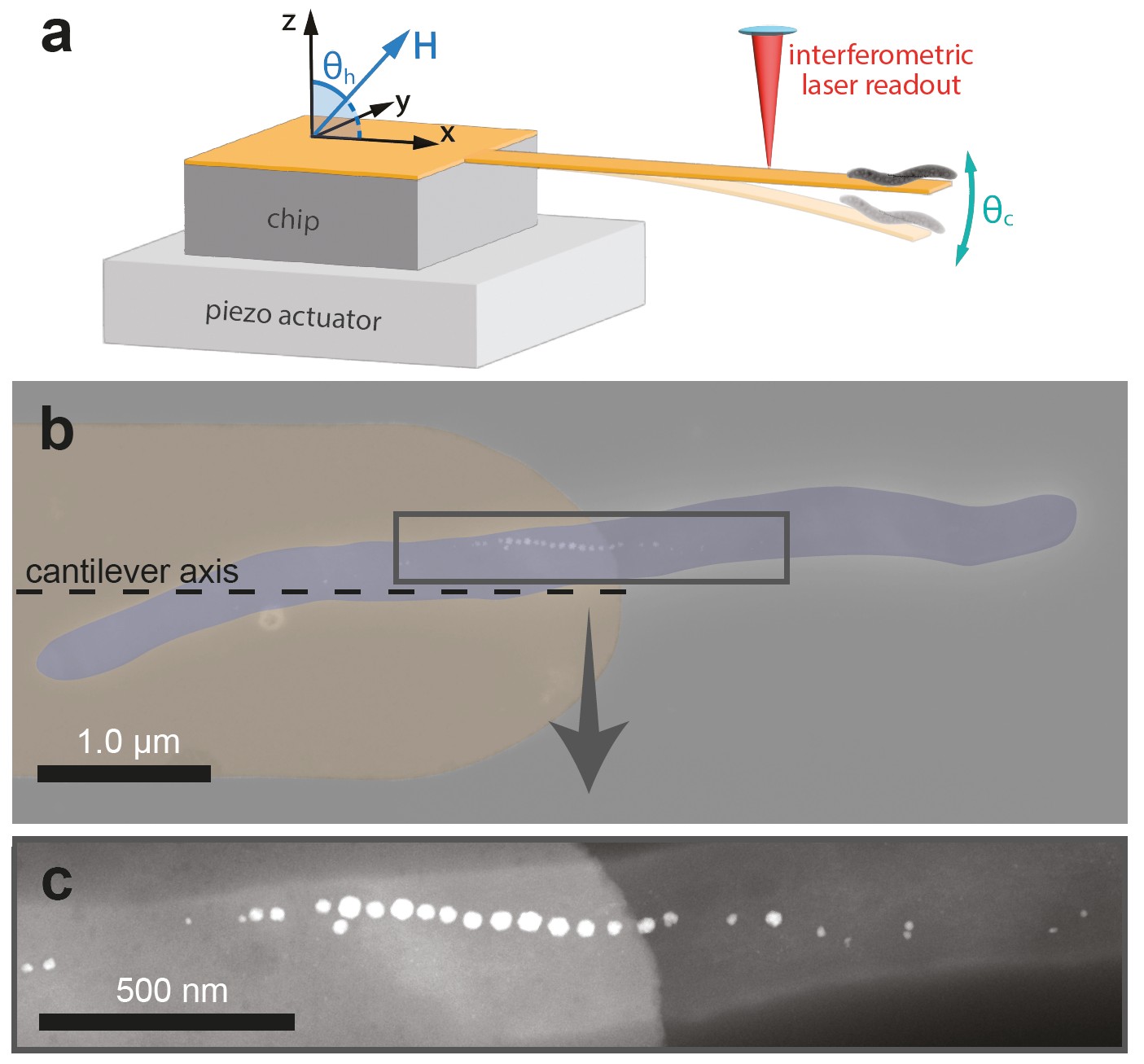}
    \caption{\textbf{Measurement setup and bacterium-on-cantilever probe.} \textbf{a} Sketch of a cantilever with a single bacterium attached at the tip and definition of the coordinate axes. \textbf{b} False color high-angle annular dark-field scanning transmission electron microscopy image showing the relative orientation of the magnetosome chain with respect to the cantilever axis and \textbf{c} a close-up view of the magnetosome chain.}
    \label{fig:fig01}
\end{figure}

Cells of \textit{M. gryph.} DSM 6361 were grown under anaerobic conditions in sealed Hungate tubes containing 10 ml of Flask Standard Medium under a headspace of N$_2$, as described before~\cite{heyen_growth_2003, li_periplasmic_2012}.
The cells are then fixed with formaldehyde. Next, a droplet of solution containing the cells is placed on the surface of a polytetrafluoroethylene sheet. A small permanent magnet placed under the sheet helps to retain bacteria, which contain magnetosomes. After 10 minutes, the droplet is dried with a gentle flow of compressed air. A micromanipulator system (Narishige) is used to pick up an individual bacterium from the surface, and transfer it to the free end of the ultra-soft cantilever. 
The bacterium is attached to the apex of the cantilever for magnetic characterization as shown in Fig.~\ref{fig:fig01}b, oriented roughly with the cantilever axis using the micromanipulator, and fixed in place with UV glue (Thorlabs).
\section{Measurements}
%
%
We perform two types of DCM measurements just below room temperature at $T\approx\SI{284}{\kelvin}$. In the first, shown in Fig.~\ref{fig:fig02}, we monitor $\Delta\!f$ as a function of the direction $\theta_h$ of a large field of constant magnitude $\mu_0 H = 3.5$~T applied in the plane perpendicular to the cantilever's rotation axis ($xz$-plane). In a magnetic field $H$ large enough to saturate the system, as in this case, the maxima and minima of $\Delta\!f (\theta_h)$ indicate the directions of the easy and hard axes, respectively~\cite{gross_dynamic_2016,gross_magnetic_2021}. The data plotted in Fig.~\ref{fig:fig02} show the signatures of uniaxial anisotropy with the maximum and minimum separated by \SI{90}{\degree} and the maximum indicating the direction of the easy axis, which roughly coincides with the axis of the magnetosome chain.

In the second type of measurement, shown in Fig.~\ref{fig:fig03}, we measure $\Delta\!f$ as we sweep $H$ along a fixed direction from large positive to large negative values. Analysis of the resulting $\Delta\!f (H)$ hysteresis measurements allows for the extraction of information on the remanent magnetic moment as well as the system's magnetic switching behavior and the progression of magnetic configurations present as the system reverses. Measurements taken for $\theta_h = \SI{90}{\degree}$ (along the axis of the magnetosome chain), shown in Fig.~\ref{fig:fig03}a and b, and for $\theta_h = \SI{0}{\degree}$, shown in Fig.~\ref{fig:fig03}c and d, show the characteristic response of a magnetic system with uniaxial anisotropy with $\mathbf{H}$ applied along its easy and hard axes, respectively~\cite{gross_dynamic_2016}.  The sharp discontinuity seen in Fig.~\ref{fig:fig03}b is the signature of magnetic switching in reverse applied field allowing us to deduce a coercive field of around 20~mT for the chain.

Upon completion of the DCM measurements, the cantilever is transferred to a grid for subsequent high-resolution (HR)TEM, high-angle annular dark-field scanning transmission electron microscopy (HAADF-STEM) imaging, and TEM tomography. During this process, the cantilever is broken off of its base at its clamping point, making further DCM measurements impossible. Individual magnetite magnetosomes, which are only surrounded by biological tissue, are clearly resolved in the HAADF-STEM images, as shown in Fig.~\ref{fig:fig01}b. This image shows the individual bacterium under investigation, which is still attached to the end of the SiN cantilever. The brightest contrast in the image corresponds to the projection of each magnetosome on the $xy$-plane. The image reveals the alignment of the magnetosome chain with the cantilever's long axis. Using TEM tomography in combination with a deep learning algorithm allows the construction of a realistic 3D model of the particles, revealing their position in 3D space and their morphology.
A list of the coordinates and volumes of each magnetosome is given in Ref.~\onlinecite{SOM}, as well as details on the 3D model construction process. 

HRTEM images of each individual magnetosome and their Fourier transforms also provide information on their crystallinity and orientation. Crystal planes visible in the images show that at least 25 of the 29 magnetosomes are in a crystalline state. We do not find correlations between the orientation of these planes and either the axis of the chain or the orientation of neighboring magnetosomes~\cite{SOM}. Experimental circumstances prevented the extraction of the exact crystalline orientation of each magnetosome, which would have provided the orientation of each particle's magneto-crystalline anisotropy axes. 

\section{Analysis and Discussion}
\begin{figure}[t]
    \centering
    \includegraphics[width=1.0\linewidth]{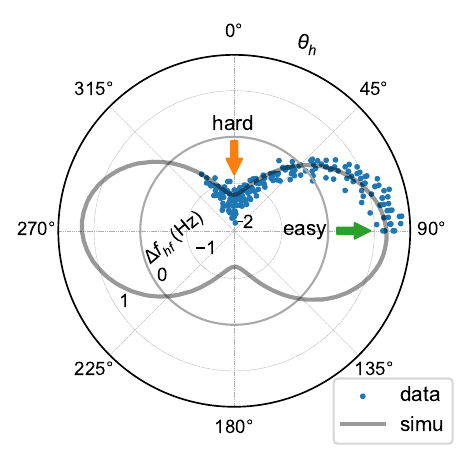}
    \caption{\textbf{DCM field rotation measurement and simulation.} $\Delta\!f(\theta_h)$ measured at $\mu{}_0H=3.5$\,T is plotted as blue points. Orange and green arrows indicate the hard and easy axes, respectively, as derived from simulation, shown in transparent grey.}
    \label{fig:fig02}
\end{figure}
DCM data plotted in Fig.~\ref{fig:fig02} indicate the presence of an effective anisotropy along the chain of magnetosomes of uniaxial character. 
A detailed consideration is necessary to understand the role the involved magnetic anisotropies play. To do so, we construct a three-dimensional, micromagnetic model of the magnetosome chain using the software package \textit{Mumax3}~\cite{vansteenkiste_design_2014, exl_labontes_2014}. The software solves the space and time dependent Landau-Lifshitz-Gilbert equation using a finite-difference approach, i.e.\ we discretize space to \SI{1.2}{\nano\meter}, and determine the relaxed state of each voxel's magnetic moment in an effective magnetic field. This effective field contains all relevant interactions and anisotropy contributions: exchange interactions, dipolar interactions, cubic magneto-crystalline anisotropy of the individual particles, and an externally applied magnetic field. Ultimately, this approach allows us to precisely model the magnetic behavior of the magnetosome chain from knowledge of the position, shape, and orientation of each magnetosome as measured by TEM. Using material parameters from the literature~\cite{o_conbhui_merrill_2018}, the model determines the equilibrium magnetic state of the chain for a set external magnetic fields. By including the geometry of the DCM experiment and the properties of the cantilever, we calculate $\Delta\!f$ and use it to iteratively adjust the input parameters, optimizing the agreement between simulated and measured $\Delta\!f$. In this way, features in hysteresis curves of $\Delta\!f (H)$ such as jumps, dips, or humps can be associated with events such as magnetic switching or collective rotation of magnetic moments.

\begin{figure*}[t]
    \centering
\includegraphics[width=1.0\linewidth]{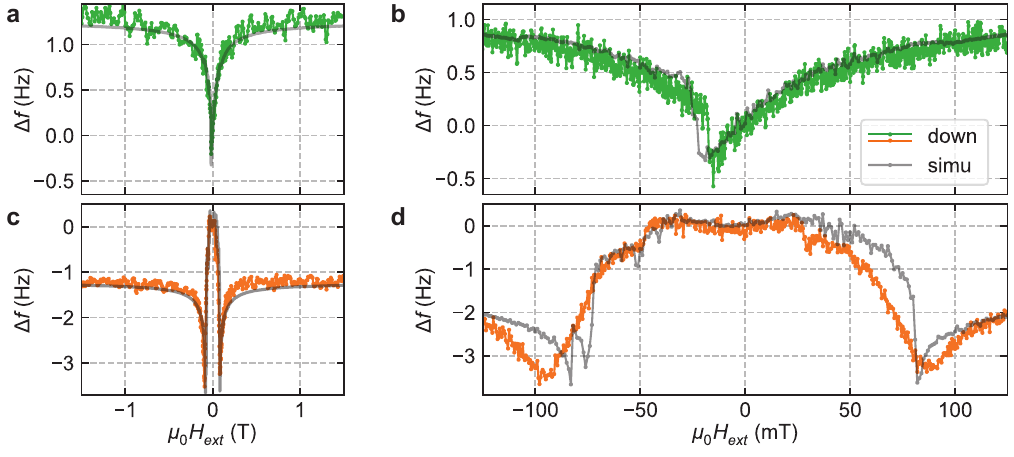}
    \caption{\textbf{DCM hysteresis as a function of applied field.} \textbf{a} and \textbf{b} Measured $\Delta\!f (H)$ with \textbf{H} parallel to the magnetic easy axis plotted in green for a wide and narrow field range, respectively. \textbf{c} and \textbf{d} Measured $\Delta\!f (H)$ with \textbf{H} perpendicular to the magnetic easy axis plotted in orange, for a wide and narrow field range, respectively. In all plots, the applied field $\mu_0 H_{ext}$ is swept from large positive values, through zero, to large negative values. Best-fit simulations are plotted in transparent grey. 
    }
    \label{fig:fig03}
\end{figure*}

Our lack of knowledge of the orientation of the crystalline anisotropy axes of the individual magnetosomes forces us to fall back on a `best guess' strategy in order to construct a realistic model: We simulate 1000 iterations of the high-field $\Delta\!f (\theta_h)$ measurement shown in Fig.~\ref{fig:fig02} with randomly oriented crystalline anisotropy axes for each magnetosome. The parameters from the 20 best matches are then used to simulate the $\Delta\!f (H)$ hysteresis measurements shown in Fig.~\ref{fig:fig03} for a number of magnetic field orientations close to the experimental orientations. Further details on the simulations can be found in Ref.~\onlinecite{SOM}.

One set of parameters is found to provide the best match to the measurements, coinciding with both the experimental $\Delta\!f (\theta_h)$ and $\Delta\!f (H)$ data within the measurement noise, as shown in Figs.~\ref{fig:fig02} and \ref{fig:fig03}. The model allows the extraction of an effective uniaxial anisotropy constant for the chain as a whole, containing all contributions from the included anisotropies, $K_u =$~\SI{12.5}{\kilo\joule/\meter\cubed}. This is in excellent agreement with the value obtained in Ref.~\cite{orue_configuration_2018} for a chain of non-interacting magnetic dipoles. Note, however, that the model therein incorporates cubic crystalline anisotropy with one axis aligned to the chain axis for all magnetosomes, deviating from the model at hand.
Typically, in literature~\cite{faivre_magnetotactic_2008}, only a minimal magnetic moment for effective magnetotaxis is given. The effective anisotropy of the magnetosome chain is, however, an equally important quantity, because it quantifies the stability of the magnetic moment along the chain axis against perturbation.

It is interesting to note that most of the 1000 simulated high-field rotation curves do not show the characteristic shape resulting from simple uniaxial magnetic anisotropy along the chain direction, as measured for this bacterium in Fig.~\ref{fig:fig02}. Rather, they show a more complex angular dependence with signatures of the shape and randomly oriented crystalline anisotropies of the individual magnetosomes, as shown in Ref.~\onlinecite{SOM}. This finding highlights that anisotropies of the individual magnetosomes should not be neglected and suggests that the absence of features due to these anisotropies in the curve in Fig.~\ref{fig:fig02} could be a coincidence due to the specific orientations of the magnetosomes in this chain, or the result of an external process, such as e.g.\ alignment of the magnetosomes during their growth in the \textit{M. gryphiswaldense} cells. 
Nevertheless, even in iterations, in which the orientations of the individual magnetosomes result in more complex $\Delta\!f(\theta_h)$, the dominant easy axis is always set by the direction of the magnetosome chain. The fact that the chain axis sets the global energetic minimum of the magnetic anisotropy energy, despite the competing anisotropies of individual magnetosomes, ultimately ensures that every bacterium's body aligns along magnetic field lines.

Some previous studies report the alignment of a $\left\langle 111 \right\rangle$-type axis of individual magnetosomes with the chain axis for several different types of magnetotactic bacteria~\cite{gandia_elucidating_2020,orue_configuration_2018,kornig_magnetite_2014,mann_structure_1984}, including \textit{M. gryph.} We tested the consistency of our measurements with such an alignment, by simulating a number of random orientations in which each magnetosome had a $\left\langle 111 \right\rangle$-type axis pinned along the chain direction. Under this constraint, no match to the measurements was found, with an exemplary result shown in Ref.~\onlinecite{SOM}. In fact, averaged over all magnetosomes, the simulation with the best match has an angle between the chain axis and the nearest $\left\langle 111 \right\rangle$-type axis of \SI{23}{\degree} with a standard deviation of \SI{12}{\degree}. We therefore conclude that no specific alignment of the $\left\langle 111 \right\rangle$ axes with the chain axis is present in the measured bacterium.

%
Measurements of the hysteresis of $\Delta\!f (H)$, shown in Fig.~\ref{fig:fig03}, are taken with $\mathbf{H}$ aligned along the magnetic easy and hard directions, as determined by the measurements shown in Fig.~\ref{fig:fig02}. 
In the limit of low fields, the component of the remanent magnetic moment parallel to the external magnetic field applied during the hysteresis cycle  can be directly determined from $\Delta f(H)$ with the knowledge of the cantilever properties~\cite{gross_dynamic_2016}.
We find $M_x \approx (1.84\pm 0.54)\cdot10^{-16}$~\si{\ampere\meter\squared} after $\mathbf{H}$ is applied parallel to the chain axis,  which is closely reproduced  by our best-scenario micromagnetic model with $M_x \approx 1.73\cdot10^{-16}$~\si{\ampere\meter\squared}, for which the calculated $\Delta\!f (H)$ is also shown in Fig.~\ref{fig:fig03}. The saturation magnetic moment within the model is $2.0\cdot10^{-16}$~\si{\ampere\meter\squared}. Therefore, almost the full moment is preserved along $x$ in remanence after $\mathbf{H}$ is applied parallel to the chain axis.
After the application of perpendicular fields, $M_z$ nearly vanishes: $M_z \approx(0.26\pm 0.14)\cdot10^{-16}$~\si{\ampere\meter\squared}. Note, however, that this measurement does not determine the remanent magnetic moment along the $x$ or $y$ directions. These values are consistent with the orientation of the effective anisotropy along the chain axis. Further, the magnitude of the saturation moment is in very good agreement with previous measurements done via optical microscopy~\cite{reufer_switching_2014, zahn_measurement_2017} and DCM~\cite{gysin_magnetic_2011}.

\begin{figure}[t]
    \centering
    \includegraphics[width=1.0\linewidth]{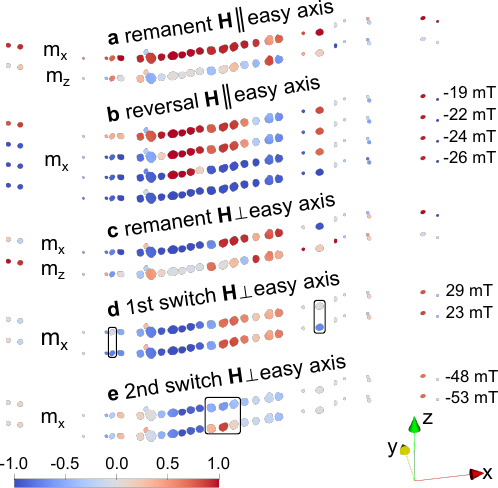}
    \caption{\textbf{Visualization of simulated magnetic states of the magnetosome chain.} The color bar applies to the component of the magnetization as indicated on the left of each of the visualisations.}
    \label{fig:fig04}
\end{figure}

Further insight can be deduced from a detailed analysis of the measured and simulated $\Delta\!f (H)$ shown in Fig.~\ref{fig:fig03}. In the measurement with the field applied along the easy-axis, shown in Fig.~\ref{fig:fig03}b, $\Delta\!f(H)$ monotonically decreases from high fields down to zero. In a reverse field of -\SI{17}{\milli\tesla}, a jump occurs and thereafter the curve progresses symmetrically as for positive fields. Such behavior and the V-shape of the curve can be explained via a single macro-spin approximation~\cite{gross_dynamic_2016}: all magnetic moments are aligned with the external field, coinciding with the easy-axis of the effective magnetic anisotropy and, shortly after field reversal, collectively flip their direction in a single switching event. Simulations reveal that this picture conceals a more complex behavior: as the applied field is swept from saturation down towards zero, some magnetosomes reorient the alignment of their magnetic moments from coinciding with the chain axis to coinciding with one of their local magnetic easy axes. In the given simulation this takes place in form of a smooth rotation with decreasing magnetic field. To illustrate the result of these rotations, Fig.~\ref{fig:fig04}a shows $m_x$ and $m_z$ of the remanent state of the magnetosome chain, where only the main part of the chain maintains close-to full alignment of the magnetic moments with the overall magnetic easy axis. On average, the remanent magnetic moment of a magnetosome and the overall magnetic easy axis span an angle of about \SI{34}{\degree} with \SI{29}{\degree} standard deviation. This qualitatively confirms a remanent magnetic moment tilted away from the overall easy anisotropy axis as observed in Ref.~\cite{orue_configuration_2018}. Although the noise level of the measurement does not allow for the confirmation of related switching events or rotations of magnetic moments in the measured $\Delta\!f (H)$, the slightly reduced remanent magnetic moment compared to the saturation moment is their direct experimental consequence. The simulations show that further sweeping $H$ in the reverse field direction results in the full magnetic reversal of the chain via switching events of individual or a few magnetosomes, as illustrated in Fig.~\ref{fig:fig04}b.

When $H$ is applied along the magnetic hard-axis, the form of the measured $\Delta\!f(H)$ is consistent with a gradual rotation of all magnetic moments from alignment with $\mathbf{H}$ towards the easy-axis for decreasing $|H|$ (compare Fig.~\ref{fig:fig03}c to Fig.~9 in Ref.~\onlinecite{gross_dynamic_2016}). The beginning of this rotation is marked by the strong dips around $\pm$\SI{100}{\milli\tesla}. Deviations from this simple behavior appear in the discontinuities at \SI{28}{\milli\tesla} and -\SI{48}{\milli\tesla} in the measured $\Delta\!f (H)$. The micromagnetic simulations, which reproduce these distinctive features at nearly the same fields, suggest that they originate in an abrupt reorientation of the magnetization of individual or groups of magnetosomes, cf. Fig.~\ref{fig:fig04}d and e. The discontinuity in d is connected to two individual magnetosomes in the chain reorienting its magnetization, while it corresponds in e to a switch of the magnetization of a group of three magnetosomes aligned with $-\hat{x}$ towards $\hat{x}$. More details can be found in the supplementary videos~\cite{SOM}. Further switching events are either hidden in the noise of the measurement or the progression of the magnetic moments takes place slightly differently than in our best-guess simulation.

The latter further suggest a specific remanent state, as shown in Fig.~\ref{fig:fig04}c. In this state, the magnetic moments on the left half of the chain orient their moments towards the left ($-\hat{x}$), while those on the right orient towards the right ($+\hat{x}$). Together the moments compensate each other, leading to a very low remanent moment compared to the case with $\mathbf{H}$ applied along the easy-axis. Note, however, that in other simulations, which match similarly well with experiment, this state of low remanent moment is not evident, but a state rather close to what is shown in Fig.~\ref{fig:fig04}a is observed.

As the progressions of the chain's magnetic configuration for the two different orientations of the external field show, the total magnetic moment at remanence depends strongly on the magnetic history of the chain. This dependence is a result of the relatively weak effective anisotropy, which fails to stabilize the magnetic moment in direction of the easy axis against the application of perpendicular fields larger than several tens of millitesla. Nevertheless, because in its natural habitat \textit{M. gryph.} likely never experiences magnetic fields of this magnitude, this effective anisotropy is sufficient to stabilize its magnetic moment in direction of the magnetosome chain and ensure magnetotaxis. Knowledge of the magnetic behavior for field magnitudes beyond earth's magnetic field may become decisive for applications of magnetotactic bacteria such as biological micro-robots, in which stronger fields are typically used.

\section{Conclusion}
Via ultrasensitive torque magnetometry, combined with transmission electron microscopy and micromagnetic simulations, we determine the remanent magnetic moment of $(1.84\pm0.54)\cdot10^{-16}$~\si{\ampere\meter\squared} and a saturation moment of $2.0\cdot10^{-16}$~\si{\ampere\meter\squared} of an individual \textit{M. gryph.} cell. Furthermore, we determine an effective easy-axis anisotropy of \SI{12.5}{\kilo\joule/\meter\cubed} aligned along the bacterium's chain of magnetosomes. Analysis of the magnetic hysteresis shows that this anisotropy is strong enough to stabilize the remanent magnetic moment in Earth's magnetic field. Future experiments, making use of TEM tomography and diffraction to reveal the full morphology and crystal structure of the magnetosomes~\cite{orue_configuration_2018}, may allow for higher precision determination of the exact magnetic states of a single bacterium. Improved cantilever sensors could also enable the resolution of the switching of individual magnetosomes, further deepening our understanding of the magnetic reversal process. DCM Measurements of many more individual bacteria would also help to determine the homogeneity of saturation moment, remanent moment, and anisotropy across a population of bacteria.
\section{Data availability}
The data supporting the findings of this study are available on the Zenodo repository at https://doi.org/10.5281/zenodo.10730902.
\section{Acknowledgments}
The authors thank Sascha Martin and his team in the machine shop of the Department of Physics at the University of Basel for help building the measurement system, Urs Gysin, the Nano Imaging Lab of the Swiss Nanoscience Institute and the BioEM lab of the University of Basel for their support in taking TEM data. They acknowledge support from the Canton Aargau and the Swiss Nanoscience Institute's PhD program (Project P2107). Calculations were performed at sciCORE (http://scicore.unibas.ch/) scientific computing center at University of Basel.

%
%
%

\begin{thebibliography}{63}%
\makeatletter
\providecommand \@ifxundefined [1]{%
 \@ifx{#1\undefined}
}%
\providecommand \@ifnum [1]{%
 \ifnum #1\expandafter \@firstoftwo
 \else \expandafter \@secondoftwo
 \fi
}%
\providecommand \@ifx [1]{%
 \ifx #1\expandafter \@firstoftwo
 \else \expandafter \@secondoftwo
 \fi
}%
\providecommand \natexlab [1]{#1}%
\providecommand \enquote  [1]{``#1''}%
\providecommand \bibnamefont  [1]{#1}%
\providecommand \bibfnamefont [1]{#1}%
\providecommand \citenamefont [1]{#1}%
\providecommand \href@noop [0]{\@secondoftwo}%
\providecommand \href [0]{\begingroup \@sanitize@url \@href}%
\providecommand \@href[1]{\@@startlink{#1}\@@href}%
\providecommand \@@href[1]{\endgroup#1\@@endlink}%
\providecommand \@sanitize@url [0]{\catcode `\\12\catcode `\$12\catcode
  `\&12\catcode `\#12\catcode `\^12\catcode `\_12\catcode `\%12\relax}%
\providecommand \@@startlink[1]{}%
\providecommand \@@endlink[0]{}%
\providecommand \url  [0]{\begingroup\@sanitize@url \@url }%
\providecommand \@url [1]{\endgroup\@href {#1}{\urlprefix }}%
\providecommand \urlprefix  [0]{URL }%
\providecommand \Eprint [0]{\href }%
\providecommand \doibase [0]{https://doi.org/}%
\providecommand \selectlanguage [0]{\@gobble}%
\providecommand \bibinfo  [0]{\@secondoftwo}%
\providecommand \bibfield  [0]{\@secondoftwo}%
\providecommand \translation [1]{[#1]}%
\providecommand \BibitemOpen [0]{}%
\providecommand \bibitemStop [0]{}%
\providecommand \bibitemNoStop [0]{.\EOS\space}%
\providecommand \EOS [0]{\spacefactor3000\relax}%
\providecommand \BibitemShut  [1]{\csname bibitem#1\endcsname}%
\let\auto@bib@innerbib\@empty
\bibitem [{\citenamefont {Uebe}\ and\ \citenamefont
  {Schüler}(2016)}]{uebe_magnetosome_2016}%
  \BibitemOpen
  \bibfield  {author} {\bibinfo {author} {\bibfnamefont {R.}~\bibnamefont
  {Uebe}}\ and\ \bibinfo {author} {\bibfnamefont {D.}~\bibnamefont
  {Schüler}},\ }\bibfield  {title} {\bibinfo {title} {Magnetosome biogenesis
  in magnetotactic bacteria},\ }\href {https://doi.org/10.1038/nrmicro.2016.99}
  {\bibfield  {journal} {\bibinfo  {journal} {Nature Reviews Microbiology}\
  }\textbf {\bibinfo {volume} {14}},\ \bibinfo {pages} {621} (\bibinfo {year}
  {2016})}\BibitemShut {NoStop}%
\bibitem [{\citenamefont {Faivre}\ and\ \citenamefont
  {Schüler}(2008)}]{faivre_magnetotactic_2008}%
  \BibitemOpen
  \bibfield  {author} {\bibinfo {author} {\bibfnamefont {D.}~\bibnamefont
  {Faivre}}\ and\ \bibinfo {author} {\bibfnamefont {D.}~\bibnamefont
  {Schüler}},\ }\bibfield  {title} {\bibinfo {title} {Magnetotactic {Bacteria}
  and {Magnetosomes}},\ }\href {https://doi.org/10.1021/cr078258w} {\bibfield
  {journal} {\bibinfo  {journal} {Chemical Reviews}\ }\textbf {\bibinfo
  {volume} {108}},\ \bibinfo {pages} {4875} (\bibinfo {year}
  {2008})}\BibitemShut {NoStop}%
\bibitem [{\citenamefont {Lefèvre}\ and\ \citenamefont
  {Bazylinski}(2013)}]{lefevre_ecology_2013}%
  \BibitemOpen
  \bibfield  {author} {\bibinfo {author} {\bibfnamefont {C.~T.}\ \bibnamefont
  {Lefèvre}}\ and\ \bibinfo {author} {\bibfnamefont {D.~A.}\ \bibnamefont
  {Bazylinski}},\ }\bibfield  {title} {\bibinfo {title} {Ecology, {Diversity},
  and {Evolution} of {Magnetotactic} {Bacteria}},\ }\href
  {https://doi.org/10.1128/mmbr.00021-13} {\bibfield  {journal} {\bibinfo
  {journal} {Microbiology and Molecular Biology Reviews}\ }\textbf {\bibinfo
  {volume} {77}},\ \bibinfo {pages} {497} (\bibinfo {year} {2013})}\BibitemShut
  {NoStop}%
\bibitem [{\citenamefont {Popp}\ \emph {et~al.}(2014)\citenamefont {Popp},
  \citenamefont {Armitage},\ and\ \citenamefont
  {Schüler}}]{popp_polarity_2014}%
  \BibitemOpen
  \bibfield  {author} {\bibinfo {author} {\bibfnamefont {F.}~\bibnamefont
  {Popp}}, \bibinfo {author} {\bibfnamefont {J.~P.}\ \bibnamefont {Armitage}},\
  and\ \bibinfo {author} {\bibfnamefont {D.}~\bibnamefont {Schüler}},\
  }\bibfield  {title} {\bibinfo {title} {Polarity of bacterial magnetotaxis is
  controlled by aerotaxis through a common sensory pathway},\ }\href
  {https://doi.org/10.1038/ncomms6398} {\bibfield  {journal} {\bibinfo
  {journal} {Nature Communications}\ }\textbf {\bibinfo {volume} {5}},\
  \bibinfo {pages} {5398} (\bibinfo {year} {2014})}\BibitemShut {NoStop}%
\bibitem [{\citenamefont {Frankel}\ \emph {et~al.}(2018)\citenamefont
  {Frankel}, \citenamefont {Bazylinski}, \citenamefont {Johnson},\ and\
  \citenamefont {Taylor}}]{frankel_magneto-aerotaxis_1997}%
  \BibitemOpen
  \bibfield  {author} {\bibinfo {author} {\bibfnamefont {R.~B.}\ \bibnamefont
  {Frankel}}, \bibinfo {author} {\bibfnamefont {D.~A.}\ \bibnamefont
  {Bazylinski}}, \bibinfo {author} {\bibfnamefont {M.~S.}\ \bibnamefont
  {Johnson}},\ and\ \bibinfo {author} {\bibfnamefont {B.~L.}\ \bibnamefont
  {Taylor}},\ }\bibfield  {title} {\bibinfo {title} {Magneto-aerotaxis in
  marine coccoid bacteria},\ }\href@noop {} {\bibfield  {journal} {\bibinfo
  {journal} {Biophysical Journal}\ }\textbf {\bibinfo {volume} {73}},\ \bibinfo
  {pages} {994} (\bibinfo {year} {2018})}\BibitemShut {NoStop}%
\bibitem [{\citenamefont {Schüler}(2002)}]{schuler_biomineralization_2002}%
  \BibitemOpen
  \bibfield  {author} {\bibinfo {author} {\bibfnamefont {D.}~\bibnamefont
  {Schüler}},\ }\bibfield  {title} {\bibinfo {title} {The biomineralization of
  magnetosomes in \textit{Magnetospirillum gryphiswaldense}},\ }\href@noop {}
  {\bibfield  {journal} {\bibinfo  {journal} {International Microbiology}\
  }\textbf {\bibinfo {volume} {5}},\ \bibinfo {pages} {209} (\bibinfo {year}
  {2002})}\BibitemShut {NoStop}%
\bibitem [{\citenamefont {Alphandéry}\ \emph {et~al.}(2013)\citenamefont
  {Alphandéry}, \citenamefont {Chebbi}, \citenamefont {Guyot},\ and\
  \citenamefont {Durand-Dubief}}]{alphandery_use_2013}%
  \BibitemOpen
  \bibfield  {author} {\bibinfo {author} {\bibfnamefont {E.}~\bibnamefont
  {Alphandéry}}, \bibinfo {author} {\bibfnamefont {I.}~\bibnamefont {Chebbi}},
  \bibinfo {author} {\bibfnamefont {F.}~\bibnamefont {Guyot}},\ and\ \bibinfo
  {author} {\bibfnamefont {M.}~\bibnamefont {Durand-Dubief}},\ }\bibfield
  {title} {\bibinfo {title} {Use of bacterial magnetosomes in the magnetic
  hyperthermia treatment of tumours: {A} review},\ }\href
  {https://doi.org/10.3109/02656736.2013.821527} {\bibfield  {journal}
  {\bibinfo  {journal} {International Journal of Hyperthermia}\ }\textbf
  {\bibinfo {volume} {29}},\ \bibinfo {pages} {801} (\bibinfo {year}
  {2013})}\BibitemShut {NoStop}%
\bibitem [{\citenamefont {Serantes}\ \emph {et~al.}(2014)\citenamefont
  {Serantes}, \citenamefont {Simeonidis}, \citenamefont {Angelakeris},
  \citenamefont {Chubykalo-Fesenko}, \citenamefont {Marciello}, \citenamefont
  {Morales}, \citenamefont {Baldomir},\ and\ \citenamefont
  {Martinez-Boubeta}}]{serantes_multiplying_2014}%
  \BibitemOpen
  \bibfield  {author} {\bibinfo {author} {\bibfnamefont {D.}~\bibnamefont
  {Serantes}}, \bibinfo {author} {\bibfnamefont {K.}~\bibnamefont
  {Simeonidis}}, \bibinfo {author} {\bibfnamefont {M.}~\bibnamefont
  {Angelakeris}}, \bibinfo {author} {\bibfnamefont {O.}~\bibnamefont
  {Chubykalo-Fesenko}}, \bibinfo {author} {\bibfnamefont {M.}~\bibnamefont
  {Marciello}}, \bibinfo {author} {\bibfnamefont {M.~d.~P.}\ \bibnamefont
  {Morales}}, \bibinfo {author} {\bibfnamefont {D.}~\bibnamefont {Baldomir}},\
  and\ \bibinfo {author} {\bibfnamefont {C.}~\bibnamefont {Martinez-Boubeta}},\
  }\bibfield  {title} {\bibinfo {title} {Multiplying {Magnetic} {Hyperthermia}
  {Response} by {Nanoparticle} {Assembling}},\ }\href
  {https://doi.org/10.1021/jp410717m} {\bibfield  {journal} {\bibinfo
  {journal} {The Journal of Physical Chemistry C}\ }\textbf {\bibinfo {volume}
  {118}},\ \bibinfo {pages} {5927} (\bibinfo {year} {2014})}\BibitemShut
  {NoStop}%
\bibitem [{\citenamefont {Felfoul}\ \emph {et~al.}(2016)\citenamefont
  {Felfoul}, \citenamefont {Mohammadi}, \citenamefont {Taherkhani},
  \citenamefont {de~Lanauze}, \citenamefont {Zhong~Xu}, \citenamefont {Loghin},
  \citenamefont {Essa}, \citenamefont {Jancik}, \citenamefont {Houle},
  \citenamefont {Lafleur}, \citenamefont {Gaboury}, \citenamefont {Tabrizian},
  \citenamefont {Kaou}, \citenamefont {Atkin}, \citenamefont {Vuong},
  \citenamefont {Batist}, \citenamefont {Beauchemin}, \citenamefont
  {Radzioch},\ and\ \citenamefont {Martel}}]{felfoul_magneto-aerotactic_2016}%
  \BibitemOpen
  \bibfield  {author} {\bibinfo {author} {\bibfnamefont {O.}~\bibnamefont
  {Felfoul}}, \bibinfo {author} {\bibfnamefont {M.}~\bibnamefont {Mohammadi}},
  \bibinfo {author} {\bibfnamefont {S.}~\bibnamefont {Taherkhani}}, \bibinfo
  {author} {\bibfnamefont {D.}~\bibnamefont {de~Lanauze}}, \bibinfo {author}
  {\bibfnamefont {Y.}~\bibnamefont {Zhong~Xu}}, \bibinfo {author}
  {\bibfnamefont {D.}~\bibnamefont {Loghin}}, \bibinfo {author} {\bibfnamefont
  {S.}~\bibnamefont {Essa}}, \bibinfo {author} {\bibfnamefont {S.}~\bibnamefont
  {Jancik}}, \bibinfo {author} {\bibfnamefont {D.}~\bibnamefont {Houle}},
  \bibinfo {author} {\bibfnamefont {M.}~\bibnamefont {Lafleur}}, \bibinfo
  {author} {\bibfnamefont {L.}~\bibnamefont {Gaboury}}, \bibinfo {author}
  {\bibfnamefont {M.}~\bibnamefont {Tabrizian}}, \bibinfo {author}
  {\bibfnamefont {N.}~\bibnamefont {Kaou}}, \bibinfo {author} {\bibfnamefont
  {M.}~\bibnamefont {Atkin}}, \bibinfo {author} {\bibfnamefont
  {T.}~\bibnamefont {Vuong}}, \bibinfo {author} {\bibfnamefont
  {G.}~\bibnamefont {Batist}}, \bibinfo {author} {\bibfnamefont
  {N.}~\bibnamefont {Beauchemin}}, \bibinfo {author} {\bibfnamefont
  {D.}~\bibnamefont {Radzioch}},\ and\ \bibinfo {author} {\bibfnamefont
  {S.}~\bibnamefont {Martel}},\ }\bibfield  {title} {\bibinfo {title}
  {Magneto-aerotactic bacteria deliver drug-containing nanoliposomes to tumour
  hypoxic regions},\ }\href {https://doi.org/10.1038/nnano.2016.137} {\bibfield
   {journal} {\bibinfo  {journal} {Nature Nanotechnology}\ }\textbf {\bibinfo
  {volume} {11}},\ \bibinfo {pages} {941} (\bibinfo {year} {2016})}\BibitemShut
  {NoStop}%
\bibitem [{\citenamefont {Ghosh}\ \emph {et~al.}(2012)\citenamefont {Ghosh},
  \citenamefont {Lee}, \citenamefont {Thomas}, \citenamefont {Kohli},
  \citenamefont {Yun}, \citenamefont {Belcher},\ and\ \citenamefont
  {Kelly}}]{ghosh_m13-templated_2012}%
  \BibitemOpen
  \bibfield  {author} {\bibinfo {author} {\bibfnamefont {D.}~\bibnamefont
  {Ghosh}}, \bibinfo {author} {\bibfnamefont {Y.}~\bibnamefont {Lee}}, \bibinfo
  {author} {\bibfnamefont {S.}~\bibnamefont {Thomas}}, \bibinfo {author}
  {\bibfnamefont {A.~G.}\ \bibnamefont {Kohli}}, \bibinfo {author}
  {\bibfnamefont {D.~S.}\ \bibnamefont {Yun}}, \bibinfo {author} {\bibfnamefont
  {A.~M.}\ \bibnamefont {Belcher}},\ and\ \bibinfo {author} {\bibfnamefont
  {K.~A.}\ \bibnamefont {Kelly}},\ }\bibfield  {title} {\bibinfo {title}
  {M13-templated magnetic nanoparticles for targeted in vivo imaging of
  prostate cancer},\ }\href {https://doi.org/10.1038/nnano.2012.146} {\bibfield
   {journal} {\bibinfo  {journal} {Nature Nanotechnology}\ }\textbf {\bibinfo
  {volume} {7}},\ \bibinfo {pages} {677} (\bibinfo {year} {2012})}\BibitemShut
  {NoStop}%
\bibitem [{\citenamefont {Gandia}\ \emph {et~al.}(2019)\citenamefont {Gandia},
  \citenamefont {Gandarias}, \citenamefont {Rodrigo}, \citenamefont
  {Robles-Garc\'ia}, \citenamefont {Das}, \citenamefont {Garaio}, \citenamefont
  {Garc\'ia}, \citenamefont {Phan}, \citenamefont {Srikanth}, \citenamefont
  {Orue}, \citenamefont {Alonso}, \citenamefont {Muela},\ and\ \citenamefont
  {Fern\'andez-Gubieda}}]{gandia_unlocking_2019}%
  \BibitemOpen
  \bibfield  {author} {\bibinfo {author} {\bibfnamefont {D.}~\bibnamefont
  {Gandia}}, \bibinfo {author} {\bibfnamefont {L.}~\bibnamefont {Gandarias}},
  \bibinfo {author} {\bibfnamefont {I.}~\bibnamefont {Rodrigo}}, \bibinfo
  {author} {\bibfnamefont {J.}~\bibnamefont {Robles-Garc\'ia}}, \bibinfo
  {author} {\bibfnamefont {R.}~\bibnamefont {Das}}, \bibinfo {author}
  {\bibfnamefont {E.}~\bibnamefont {Garaio}}, \bibinfo {author} {\bibfnamefont
  {J.~A.}\ \bibnamefont {Garc\'ia}}, \bibinfo {author} {\bibfnamefont {M.-H.}\
  \bibnamefont {Phan}}, \bibinfo {author} {\bibfnamefont {H.}~\bibnamefont
  {Srikanth}}, \bibinfo {author} {\bibfnamefont {I.}~\bibnamefont {Orue}},
  \bibinfo {author} {\bibfnamefont {J.}~\bibnamefont {Alonso}}, \bibinfo
  {author} {\bibfnamefont {A.}~\bibnamefont {Muela}},\ and\ \bibinfo {author}
  {\bibfnamefont {M.~L.}\ \bibnamefont {Fern\'andez-Gubieda}},\ }\bibfield
  {title} {\bibinfo {title} {Unlocking the {Potential} of {Magnetotactic}
  {Bacteria} as {Magnetic} {Hyperthermia} {Agents}},\ }\href
  {https://doi.org/10.1002/smll.201902626} {\bibfield  {journal} {\bibinfo
  {journal} {Small}\ }\textbf {\bibinfo {volume} {15}},\ \bibinfo {pages}
  {1902626} (\bibinfo {year} {2019})}\BibitemShut {NoStop}%
\bibitem [{\citenamefont {Kraupner}\ \emph {et~al.}(2017)\citenamefont
  {Kraupner}, \citenamefont {Eberbeck}, \citenamefont {Heinke}, \citenamefont
  {Uebe}, \citenamefont {Schüler},\ and\ \citenamefont
  {Briel}}]{kraupner_bacterial_2017}%
  \BibitemOpen
  \bibfield  {author} {\bibinfo {author} {\bibfnamefont {A.}~\bibnamefont
  {Kraupner}}, \bibinfo {author} {\bibfnamefont {D.}~\bibnamefont {Eberbeck}},
  \bibinfo {author} {\bibfnamefont {D.}~\bibnamefont {Heinke}}, \bibinfo
  {author} {\bibfnamefont {R.}~\bibnamefont {Uebe}}, \bibinfo {author}
  {\bibfnamefont {D.}~\bibnamefont {Schüler}},\ and\ \bibinfo {author}
  {\bibfnamefont {A.}~\bibnamefont {Briel}},\ }\bibfield  {title} {\bibinfo
  {title} {Bacterial magnetosomes – nature's powerful contribution to {MPI}
  tracer research},\ }\href {https://doi.org/10.1039/C7NR01530E} {\bibfield
  {journal} {\bibinfo  {journal} {Nanoscale}\ }\textbf {\bibinfo {volume}
  {9}},\ \bibinfo {pages} {5788} (\bibinfo {year} {2017})}\BibitemShut
  {NoStop}%
\bibitem [{\citenamefont {Mickoleit}\ \emph {et~al.}(2020)\citenamefont
  {Mickoleit}, \citenamefont {Lanzloth},\ and\ \citenamefont
  {Schüler}}]{mickoleit_versatile_2020}%
  \BibitemOpen
  \bibfield  {author} {\bibinfo {author} {\bibfnamefont {F.}~\bibnamefont
  {Mickoleit}}, \bibinfo {author} {\bibfnamefont {C.}~\bibnamefont
  {Lanzloth}},\ and\ \bibinfo {author} {\bibfnamefont {D.}~\bibnamefont
  {Schüler}},\ }\bibfield  {title} {\bibinfo {title} {A {Versatile} {Toolkit}
  for {Controllable} and {Highly} {Selective} {Multifunctionalization} of
  {Bacterial} {Magnetic} {Nanoparticles}},\ }\href
  {https://doi.org/10.1002/smll.201906922} {\bibfield  {journal} {\bibinfo
  {journal} {Small}\ }\textbf {\bibinfo {volume} {16}},\ \bibinfo {pages}
  {1906922} (\bibinfo {year} {2020})}\BibitemShut {NoStop}%
\bibitem [{\citenamefont {Khalil}\ and\ \citenamefont
  {Misra}(2017)}]{khalil_chapter_2017}%
  \BibitemOpen
  \bibfield  {author} {\bibinfo {author} {\bibfnamefont {I.~S.~M.}\
  \bibnamefont {Khalil}}\ and\ \bibinfo {author} {\bibfnamefont
  {S.}~\bibnamefont {Misra}},\ }\bibfield  {title} {\bibinfo {title} {Chapter 4
  - {Control} of magnetotactic bacteria},\ }in\ \href
  {https://doi.org/10.1016/B978-0-32-342993-1.00010-0} {\emph {\bibinfo
  {booktitle} {Microbiorobotics ({Second} {Edition})}}},\ \bibinfo {series and
  number} {Micro and {Nano} {Technologies}},\ \bibinfo {editor} {edited by\
  \bibinfo {editor} {\bibfnamefont {M.}~\bibnamefont {Kim}}, \bibinfo {editor}
  {\bibfnamefont {A.~A.}\ \bibnamefont {Julius}},\ and\ \bibinfo {editor}
  {\bibfnamefont {U.~K.}\ \bibnamefont {Cheang}}}\ (\bibinfo  {publisher}
  {Elsevier},\ \bibinfo {address} {Boston},\ \bibinfo {year} {2017})\ pp.\
  \bibinfo {pages} {61--79}\BibitemShut {NoStop}%
\bibitem [{\citenamefont {Mishra}\ \emph {et~al.}(2016)\citenamefont {Mishra},
  \citenamefont {Dickey}, \citenamefont {Velev},\ and\ \citenamefont
  {Tracy}}]{mishra_selective_2016}%
  \BibitemOpen
  \bibfield  {author} {\bibinfo {author} {\bibfnamefont {S.~R.}\ \bibnamefont
  {Mishra}}, \bibinfo {author} {\bibfnamefont {M.~D.}\ \bibnamefont {Dickey}},
  \bibinfo {author} {\bibfnamefont {O.~D.}\ \bibnamefont {Velev}},\ and\
  \bibinfo {author} {\bibfnamefont {J.~B.}\ \bibnamefont {Tracy}},\ }\bibfield
  {title} {\bibinfo {title} {Selective and directional actuation of elastomer
  films using chained magnetic nanoparticles},\ }\href
  {https://doi.org/10.1039/C5NR07410J} {\bibfield  {journal} {\bibinfo
  {journal} {Nanoscale}\ }\textbf {\bibinfo {volume} {8}},\ \bibinfo {pages}
  {1309} (\bibinfo {year} {2016})}\BibitemShut {NoStop}%
\bibitem [{\citenamefont {Jiang}\ \emph {et~al.}(2016)\citenamefont {Jiang},
  \citenamefont {Feng}, \citenamefont {Huang}, \citenamefont {Wu},
  \citenamefont {Su}, \citenamefont {Yang}, \citenamefont {Mai},\ and\
  \citenamefont {Jiang}}]{jiang_bioinspired_2016}%
  \BibitemOpen
  \bibfield  {author} {\bibinfo {author} {\bibfnamefont {X.}~\bibnamefont
  {Jiang}}, \bibinfo {author} {\bibfnamefont {J.}~\bibnamefont {Feng}},
  \bibinfo {author} {\bibfnamefont {L.}~\bibnamefont {Huang}}, \bibinfo
  {author} {\bibfnamefont {Y.}~\bibnamefont {Wu}}, \bibinfo {author}
  {\bibfnamefont {B.}~\bibnamefont {Su}}, \bibinfo {author} {\bibfnamefont
  {W.}~\bibnamefont {Yang}}, \bibinfo {author} {\bibfnamefont {L.}~\bibnamefont
  {Mai}},\ and\ \bibinfo {author} {\bibfnamefont {L.}~\bibnamefont {Jiang}},\
  }\bibfield  {title} {\bibinfo {title} {Bioinspired {1D} {Superparamagnetic}
  {Magnetite} {Arrays} with {Magnetic} {Field} {Perception}},\ }\href
  {https://doi.org/10.1002/adma.201601609} {\bibfield  {journal} {\bibinfo
  {journal} {Advanced Materials}\ }\textbf {\bibinfo {volume} {28}},\ \bibinfo
  {pages} {6952} (\bibinfo {year} {2016})}\BibitemShut {NoStop}%
\bibitem [{\citenamefont {Schüler}(2008)}]{schuler_genetics_2008}%
  \BibitemOpen
  \bibfield  {author} {\bibinfo {author} {\bibfnamefont {D.}~\bibnamefont
  {Schüler}},\ }\bibfield  {title} {\bibinfo {title} {Genetics and cell
  biology of magnetosome formation in magnetotactic bacteria},\ }\href
  {https://doi.org/10.1111/j.1574-6976.2008.00116.x} {\bibfield  {journal}
  {\bibinfo  {journal} {FEMS Microbiology Reviews}\ }\textbf {\bibinfo {volume}
  {32}},\ \bibinfo {pages} {654} (\bibinfo {year} {2008})}\BibitemShut
  {NoStop}%
\bibitem [{\citenamefont {Zheng}\ \emph {et~al.}(2022)\citenamefont {Zheng},
  \citenamefont {Pang}, \citenamefont {Li}, \citenamefont {Ma}, \citenamefont
  {Xu}, \citenamefont {Wen},\ and\ \citenamefont
  {Tian}}]{zheng_construction_2022}%
  \BibitemOpen
  \bibfield  {author} {\bibinfo {author} {\bibfnamefont {H.}~\bibnamefont
  {Zheng}}, \bibinfo {author} {\bibfnamefont {B.}~\bibnamefont {Pang}},
  \bibinfo {author} {\bibfnamefont {S.}~\bibnamefont {Li}}, \bibinfo {author}
  {\bibfnamefont {S.}~\bibnamefont {Ma}}, \bibinfo {author} {\bibfnamefont
  {J.}~\bibnamefont {Xu}}, \bibinfo {author} {\bibfnamefont {Y.}~\bibnamefont
  {Wen}},\ and\ \bibinfo {author} {\bibfnamefont {J.}~\bibnamefont {Tian}},\
  }\bibfield  {title} {\bibinfo {title} {Construction of {Recombinant}
  {Magnetospirillum} {Strains} for {Nitrate} {Removal} from {Wastewater}
  {Based} on {Magnetic} {Adsorption}},\ }\href
  {https://doi.org/10.3390/pr10030591} {\bibfield  {journal} {\bibinfo
  {journal} {Processes}\ }\textbf {\bibinfo {volume} {10}},\ \bibinfo {pages}
  {591} (\bibinfo {year} {2022})}\BibitemShut {NoStop}%
\bibitem [{\citenamefont {Arató}\ \emph {et~al.}(2005)\citenamefont {Arató},
  \citenamefont {Szányi}, \citenamefont {Flies}, \citenamefont {Schüler},
  \citenamefont {Frankel}, \citenamefont {Buseck},\ and\ \citenamefont
  {Pósfai}}]{arato_crystal-size_2005}%
  \BibitemOpen
  \bibfield  {author} {\bibinfo {author} {\bibfnamefont {B.}~\bibnamefont
  {Arató}}, \bibinfo {author} {\bibfnamefont {Z.}~\bibnamefont {Szányi}},
  \bibinfo {author} {\bibfnamefont {C.}~\bibnamefont {Flies}}, \bibinfo
  {author} {\bibfnamefont {D.}~\bibnamefont {Schüler}}, \bibinfo {author}
  {\bibfnamefont {R.~B.}\ \bibnamefont {Frankel}}, \bibinfo {author}
  {\bibfnamefont {P.~R.}\ \bibnamefont {Buseck}},\ and\ \bibinfo {author}
  {\bibfnamefont {M.}~\bibnamefont {Pósfai}},\ }\bibfield  {title} {\bibinfo
  {title} {Crystal-size and shape distributions of magnetite from uncultured
  magnetotactic bacteria as a potential biomarker},\ }\href
  {https://doi.org/10.2138/am.2005.1778} {\bibfield  {journal} {\bibinfo
  {journal} {American Mineralogist}\ }\textbf {\bibinfo {volume} {90}},\
  \bibinfo {pages} {1233} (\bibinfo {year} {2005})}\BibitemShut {NoStop}%
\bibitem [{\citenamefont {Thomas-Keprta}\ \emph {et~al.}(2002)\citenamefont
  {Thomas-Keprta}, \citenamefont {Clemett}, \citenamefont {Bazylinski},
  \citenamefont {Kirschvink}, \citenamefont {McKay}, \citenamefont {Wentworth},
  \citenamefont {Vali}, \citenamefont {Gibson},\ and\ \citenamefont
  {Romanek}}]{thomas-keprta_magnetofossils_2002}%
  \BibitemOpen
  \bibfield  {author} {\bibinfo {author} {\bibfnamefont {K.~L.}\ \bibnamefont
  {Thomas-Keprta}}, \bibinfo {author} {\bibfnamefont {S.~J.}\ \bibnamefont
  {Clemett}}, \bibinfo {author} {\bibfnamefont {D.~A.}\ \bibnamefont
  {Bazylinski}}, \bibinfo {author} {\bibfnamefont {J.~L.}\ \bibnamefont
  {Kirschvink}}, \bibinfo {author} {\bibfnamefont {D.~S.}\ \bibnamefont
  {McKay}}, \bibinfo {author} {\bibfnamefont {S.~J.}\ \bibnamefont
  {Wentworth}}, \bibinfo {author} {\bibfnamefont {H.}~\bibnamefont {Vali}},
  \bibinfo {author} {\bibfnamefont {E.~K.}\ \bibnamefont {Gibson}},\ and\
  \bibinfo {author} {\bibfnamefont {C.~S.}\ \bibnamefont {Romanek}},\
  }\bibfield  {title} {\bibinfo {title} {Magnetofossils from {Ancient} {Mars}:
  a {Robust} {Biosignature} in the {Martian} {Meteorite} {ALH84001}},\ }\href
  {https://doi.org/10.1128/AEM.68.8.3663-3672.2002} {\bibfield  {journal}
  {\bibinfo  {journal} {Applied and Environmental Microbiology}\ }\textbf
  {\bibinfo {volume} {68}},\ \bibinfo {pages} {3663} (\bibinfo {year}
  {2002})}\BibitemShut {NoStop}%
\bibitem [{\citenamefont {Frankel}\ and\ \citenamefont
  {Buseck}(2000)}]{frankel_magnetite_2000}%
  \BibitemOpen
  \bibfield  {author} {\bibinfo {author} {\bibfnamefont {R.~B.}\ \bibnamefont
  {Frankel}}\ and\ \bibinfo {author} {\bibfnamefont {P.~R.}\ \bibnamefont
  {Buseck}},\ }\bibfield  {title} {\bibinfo {title} {Magnetite
  biomineralization and ancient life on {Mars}},\ }\href
  {https://doi.org/10.1016/S1367-5931(99)00072-1} {\bibfield  {journal}
  {\bibinfo  {journal} {Current Opinion in Chemical Biology}\ }\textbf
  {\bibinfo {volume} {4}},\ \bibinfo {pages} {171} (\bibinfo {year}
  {2000})}\BibitemShut {NoStop}%
\bibitem [{\citenamefont {Frankel}(1984)}]{frankel_magnetic_1984}%
  \BibitemOpen
  \bibfield  {author} {\bibinfo {author} {\bibfnamefont {R.~B.}\ \bibnamefont
  {Frankel}},\ }\bibfield  {title} {\bibinfo {title} {Magnetic {Guidance} of
  {Organisms}},\ }\href {https://doi.org/10.1146/annurev.bb.13.060184.000505}
  {\bibfield  {journal} {\bibinfo  {journal} {Annual Review of Biophysics and
  Bioengineering}\ }\textbf {\bibinfo {volume} {13}},\ \bibinfo {pages} {85}
  (\bibinfo {year} {1984})}\BibitemShut {NoStop}%
\bibitem [{\citenamefont {Kalmijn}(1981)}]{kalmijn_biophysics_1981}%
  \BibitemOpen
  \bibfield  {author} {\bibinfo {author} {\bibfnamefont {A.}~\bibnamefont
  {Kalmijn}},\ }\bibfield  {title} {\bibinfo {title} {Biophysics of geomagnetic
  field detection},\ }\href {https://doi.org/10.1109/TMAG.1981.1061156}
  {\bibfield  {journal} {\bibinfo  {journal} {IEEE Transactions on Magnetics}\
  }\textbf {\bibinfo {volume} {17}},\ \bibinfo {pages} {1113} (\bibinfo {year}
  {1981})}\BibitemShut {NoStop}%
\bibitem [{\citenamefont {Zhu}\ \emph {et~al.}(2014)\citenamefont {Zhu},
  \citenamefont {Ge}, \citenamefont {Li}, \citenamefont {Wu}, \citenamefont
  {Luo}, \citenamefont {Ouyang}, \citenamefont {Tu},\ and\ \citenamefont
  {Chen}}]{zhu_angle_2014}%
  \BibitemOpen
  \bibfield  {author} {\bibinfo {author} {\bibfnamefont {X.}~\bibnamefont
  {Zhu}}, \bibinfo {author} {\bibfnamefont {X.}~\bibnamefont {Ge}}, \bibinfo
  {author} {\bibfnamefont {N.}~\bibnamefont {Li}}, \bibinfo {author}
  {\bibfnamefont {L.-F.}\ \bibnamefont {Wu}}, \bibinfo {author} {\bibfnamefont
  {C.}~\bibnamefont {Luo}}, \bibinfo {author} {\bibfnamefont {Q.}~\bibnamefont
  {Ouyang}}, \bibinfo {author} {\bibfnamefont {Y.}~\bibnamefont {Tu}},\ and\
  \bibinfo {author} {\bibfnamefont {G.}~\bibnamefont {Chen}},\ }\bibfield
  {title} {\bibinfo {title} {Angle sensing in magnetotaxis of
  {Magnetospirillum} magneticum {AMB}-1},\ }\href
  {https://doi.org/10.1039/c3ib40259b} {\bibfield  {journal} {\bibinfo
  {journal} {Integrative Biology}\ }\textbf {\bibinfo {volume} {6}},\ \bibinfo
  {pages} {706} (\bibinfo {year} {2014})}\BibitemShut {NoStop}%
\bibitem [{\citenamefont {Klumpp}\ and\ \citenamefont
  {Faivre}(2016)}]{klumpp_magnetotactic_2016}%
  \BibitemOpen
  \bibfield  {author} {\bibinfo {author} {\bibfnamefont {S.}~\bibnamefont
  {Klumpp}}\ and\ \bibinfo {author} {\bibfnamefont {D.}~\bibnamefont
  {Faivre}},\ }\bibfield  {title} {\bibinfo {title} {Magnetotactic bacteria},\
  }\href {https://doi.org/10.1140/epjst/e2016-60055-y} {\bibfield  {journal}
  {\bibinfo  {journal} {The European Physical Journal Special Topics}\ }\textbf
  {\bibinfo {volume} {225}},\ \bibinfo {pages} {2173} (\bibinfo {year}
  {2016})}\BibitemShut {NoStop}%
\bibitem [{\citenamefont {Klumpp}\ \emph {et~al.}(2019)\citenamefont {Klumpp},
  \citenamefont {Lefèvre}, \citenamefont {Bennet},\ and\ \citenamefont
  {Faivre}}]{klumpp_swimming_2019}%
  \BibitemOpen
  \bibfield  {author} {\bibinfo {author} {\bibfnamefont {S.}~\bibnamefont
  {Klumpp}}, \bibinfo {author} {\bibfnamefont {C.~T.}\ \bibnamefont
  {Lefèvre}}, \bibinfo {author} {\bibfnamefont {M.}~\bibnamefont {Bennet}},\
  and\ \bibinfo {author} {\bibfnamefont {D.}~\bibnamefont {Faivre}},\
  }\bibfield  {title} {\bibinfo {title} {Swimming with magnets: {From}
  biological organisms to synthetic devices},\ }\href
  {https://doi.org/10.1016/j.physrep.2018.10.007} {\bibfield  {journal}
  {\bibinfo  {journal} {Physics Reports}\ }\bibinfo {series} {Swimming with
  magnets: {From} biological organisms to synthetic devices},\ \textbf
  {\bibinfo {volume} {789}},\ \bibinfo {pages} {1} (\bibinfo {year}
  {2019})}\BibitemShut {NoStop}%
\bibitem [{\citenamefont {Mohammadinejad}\ \emph {et~al.}(2021)\citenamefont
  {Mohammadinejad}, \citenamefont {Faivre},\ and\ \citenamefont
  {Klumpp}}]{mohammadinejad_stokesian_2021}%
  \BibitemOpen
  \bibfield  {author} {\bibinfo {author} {\bibfnamefont {S.}~\bibnamefont
  {Mohammadinejad}}, \bibinfo {author} {\bibfnamefont {D.}~\bibnamefont
  {Faivre}},\ and\ \bibinfo {author} {\bibfnamefont {S.}~\bibnamefont
  {Klumpp}},\ }\bibfield  {title} {\bibinfo {title} {Stokesian dynamics
  simulations of a magnetotactic bacterium},\ }\href
  {https://doi.org/10.1140/epje/s10189-021-00038-5} {\bibfield  {journal}
  {\bibinfo  {journal} {The European Physical Journal E}\ }\textbf {\bibinfo
  {volume} {44}},\ \bibinfo {pages} {40} (\bibinfo {year} {2021})}\BibitemShut
  {NoStop}%
\bibitem [{\citenamefont {Codutti}\ \emph {et~al.}(2022)\citenamefont
  {Codutti}, \citenamefont {Charsooghi}, \citenamefont {Cerdá-Doñate},
  \citenamefont {Taïeb}, \citenamefont {Robinson}, \citenamefont {Faivre},\
  and\ \citenamefont {Klumpp}}]{codutti_interplay_2022}%
  \BibitemOpen
  \bibfield  {author} {\bibinfo {author} {\bibfnamefont {A.}~\bibnamefont
  {Codutti}}, \bibinfo {author} {\bibfnamefont {M.~A.}\ \bibnamefont
  {Charsooghi}}, \bibinfo {author} {\bibfnamefont {E.}~\bibnamefont
  {Cerdá-Doñate}}, \bibinfo {author} {\bibfnamefont {H.~M.}\ \bibnamefont
  {Taïeb}}, \bibinfo {author} {\bibfnamefont {T.}~\bibnamefont {Robinson}},
  \bibinfo {author} {\bibfnamefont {D.}~\bibnamefont {Faivre}},\ and\ \bibinfo
  {author} {\bibfnamefont {S.}~\bibnamefont {Klumpp}},\ }\bibfield  {title}
  {\bibinfo {title} {Interplay of surface interaction and magnetic torque in
  single-cell motion of magnetotactic bacteria in microfluidic confinement},\
  }\href {https://doi.org/10.7554/eLife.71527} {\bibfield  {journal} {\bibinfo
  {journal} {eLife}\ }\textbf {\bibinfo {volume} {11}},\ \bibinfo {pages}
  {e71527} (\bibinfo {year} {2022})}\BibitemShut {NoStop}%
\bibitem [{\citenamefont {Zahn}\ \emph {et~al.}(2017)\citenamefont {Zahn},
  \citenamefont {Keller}, \citenamefont {Toro-Nahuelpan}, \citenamefont
  {Dorscht}, \citenamefont {Gross}, \citenamefont {Laumann}, \citenamefont
  {Gekle}, \citenamefont {Zimmermann}, \citenamefont {Schüler},\ and\
  \citenamefont {Kress}}]{zahn_measurement_2017}%
  \BibitemOpen
  \bibfield  {author} {\bibinfo {author} {\bibfnamefont {C.}~\bibnamefont
  {Zahn}}, \bibinfo {author} {\bibfnamefont {S.}~\bibnamefont {Keller}},
  \bibinfo {author} {\bibfnamefont {M.}~\bibnamefont {Toro-Nahuelpan}},
  \bibinfo {author} {\bibfnamefont {P.}~\bibnamefont {Dorscht}}, \bibinfo
  {author} {\bibfnamefont {W.}~\bibnamefont {Gross}}, \bibinfo {author}
  {\bibfnamefont {M.}~\bibnamefont {Laumann}}, \bibinfo {author} {\bibfnamefont
  {S.}~\bibnamefont {Gekle}}, \bibinfo {author} {\bibfnamefont
  {W.}~\bibnamefont {Zimmermann}}, \bibinfo {author} {\bibfnamefont
  {D.}~\bibnamefont {Schüler}},\ and\ \bibinfo {author} {\bibfnamefont
  {H.}~\bibnamefont {Kress}},\ }\bibfield  {title} {\bibinfo {title}
  {Measurement of the magnetic moment of single {Magnetospirillum}
  gryphiswaldense cells by magnetic tweezers},\ }\href
  {https://doi.org/10.1038/s41598-017-03756-z} {\bibfield  {journal} {\bibinfo
  {journal} {Scientific Reports}\ }\textbf {\bibinfo {volume} {7}},\ \bibinfo
  {pages} {3558} (\bibinfo {year} {2017})}\BibitemShut {NoStop}%
\bibitem [{\citenamefont {Reufer}\ \emph {et~al.}(2014)\citenamefont {Reufer},
  \citenamefont {Besseling}, \citenamefont {Schwarz-Linek}, \citenamefont
  {Martinez}, \citenamefont {Morozov}, \citenamefont {Arlt}, \citenamefont
  {Trubitsyn}, \citenamefont {Ward},\ and\ \citenamefont
  {Poon}}]{reufer_switching_2014}%
  \BibitemOpen
  \bibfield  {author} {\bibinfo {author} {\bibfnamefont {M.}~\bibnamefont
  {Reufer}}, \bibinfo {author} {\bibfnamefont {R.}~\bibnamefont {Besseling}},
  \bibinfo {author} {\bibfnamefont {J.}~\bibnamefont {Schwarz-Linek}}, \bibinfo
  {author} {\bibfnamefont {V.~A.}\ \bibnamefont {Martinez}}, \bibinfo {author}
  {\bibfnamefont {A.~N.}\ \bibnamefont {Morozov}}, \bibinfo {author}
  {\bibfnamefont {J.}~\bibnamefont {Arlt}}, \bibinfo {author} {\bibfnamefont
  {D.}~\bibnamefont {Trubitsyn}}, \bibinfo {author} {\bibfnamefont {F.~B.}\
  \bibnamefont {Ward}},\ and\ \bibinfo {author} {\bibfnamefont {W.~C.~K.}\
  \bibnamefont {Poon}},\ }\bibfield  {title} {\bibinfo {title} {Switching of
  {Swimming} {Modes} in {Magnetospirillium} gryphiswaldense},\ }\href
  {https://doi.org/10.1016/j.bpj.2013.10.038} {\bibfield  {journal} {\bibinfo
  {journal} {Biophysical Journal}\ }\textbf {\bibinfo {volume} {106}},\
  \bibinfo {pages} {37} (\bibinfo {year} {2014})}\BibitemShut {NoStop}%
\bibitem [{\citenamefont {Le~Sage}\ \emph {et~al.}(2013)\citenamefont
  {Le~Sage}, \citenamefont {Arai}, \citenamefont {Glenn}, \citenamefont
  {DeVience}, \citenamefont {Pham}, \citenamefont {Rahn-Lee}, \citenamefont
  {Lukin}, \citenamefont {Yacoby}, \citenamefont {Komeili},\ and\ \citenamefont
  {Walsworth}}]{le_sage_optical_2013}%
  \BibitemOpen
  \bibfield  {author} {\bibinfo {author} {\bibfnamefont {D.}~\bibnamefont
  {Le~Sage}}, \bibinfo {author} {\bibfnamefont {K.}~\bibnamefont {Arai}},
  \bibinfo {author} {\bibfnamefont {D.~R.}\ \bibnamefont {Glenn}}, \bibinfo
  {author} {\bibfnamefont {S.~J.}\ \bibnamefont {DeVience}}, \bibinfo {author}
  {\bibfnamefont {L.~M.}\ \bibnamefont {Pham}}, \bibinfo {author}
  {\bibfnamefont {L.}~\bibnamefont {Rahn-Lee}}, \bibinfo {author}
  {\bibfnamefont {M.~D.}\ \bibnamefont {Lukin}}, \bibinfo {author}
  {\bibfnamefont {A.}~\bibnamefont {Yacoby}}, \bibinfo {author} {\bibfnamefont
  {A.}~\bibnamefont {Komeili}},\ and\ \bibinfo {author} {\bibfnamefont {R.~L.}\
  \bibnamefont {Walsworth}},\ }\bibfield  {title} {\bibinfo {title} {Optical
  magnetic imaging of living cells},\ }\href
  {https://doi.org/10.1038/nature12072} {\bibfield  {journal} {\bibinfo
  {journal} {Nature}\ }\textbf {\bibinfo {volume} {496}},\ \bibinfo {pages}
  {486} (\bibinfo {year} {2013})}\BibitemShut {NoStop}%
\bibitem [{\citenamefont {Nadkarni}\ \emph {et~al.}(2013)\citenamefont
  {Nadkarni}, \citenamefont {Barkley},\ and\ \citenamefont
  {Fradin}}]{nadkarni_comparison_2013}%
  \BibitemOpen
  \bibfield  {author} {\bibinfo {author} {\bibfnamefont {R.}~\bibnamefont
  {Nadkarni}}, \bibinfo {author} {\bibfnamefont {S.}~\bibnamefont {Barkley}},\
  and\ \bibinfo {author} {\bibfnamefont {C.}~\bibnamefont {Fradin}},\
  }\bibfield  {title} {\bibinfo {title} {A {Comparison} of {Methods} to
  {Measure} the {Magnetic} {Moment} of {Magnetotactic} {Bacteria} through
  {Analysis} of {Their} {Trajectories} in {External} {Magnetic} {Fields}},\
  }\href {https://doi.org/10.1371/journal.pone.0082064} {\bibfield  {journal}
  {\bibinfo  {journal} {PLOS ONE}\ }\textbf {\bibinfo {volume} {8}},\ \bibinfo
  {pages} {e82064} (\bibinfo {year} {2013})}\BibitemShut {NoStop}%
\bibitem [{\citenamefont {Bahaj}\ \emph {et~al.}(1996)\citenamefont {Bahaj},
  \citenamefont {James},\ and\ \citenamefont
  {Moeschler}}]{bahaj_alternative_1996}%
  \BibitemOpen
  \bibfield  {author} {\bibinfo {author} {\bibfnamefont {A.}~\bibnamefont
  {Bahaj}}, \bibinfo {author} {\bibfnamefont {P.}~\bibnamefont {James}},\ and\
  \bibinfo {author} {\bibfnamefont {F.}~\bibnamefont {Moeschler}},\ }\bibfield
  {title} {\bibinfo {title} {An alternative method for the estimation of the
  magnetic moment of non-spherical magnetotactic bacteria},\ }\href
  {https://doi.org/10.1109/20.539514} {\bibfield  {journal} {\bibinfo
  {journal} {IEEE Transactions on Magnetics}\ }\textbf {\bibinfo {volume}
  {32}},\ \bibinfo {pages} {5133} (\bibinfo {year} {1996})}\BibitemShut
  {NoStop}%
\bibitem [{\citenamefont {Chemla}\ \emph {et~al.}(1999)\citenamefont {Chemla},
  \citenamefont {Grossman}, \citenamefont {Lee}, \citenamefont {Clarke},
  \citenamefont {Adamkiewicz},\ and\ \citenamefont
  {Buchanan}}]{chemla_new_1999}%
  \BibitemOpen
  \bibfield  {author} {\bibinfo {author} {\bibfnamefont {Y.~R.}\ \bibnamefont
  {Chemla}}, \bibinfo {author} {\bibfnamefont {H.~L.}\ \bibnamefont
  {Grossman}}, \bibinfo {author} {\bibfnamefont {T.~S.}\ \bibnamefont {Lee}},
  \bibinfo {author} {\bibfnamefont {J.}~\bibnamefont {Clarke}}, \bibinfo
  {author} {\bibfnamefont {M.}~\bibnamefont {Adamkiewicz}},\ and\ \bibinfo
  {author} {\bibfnamefont {B.~B.}\ \bibnamefont {Buchanan}},\ }\bibfield
  {title} {\bibinfo {title} {A {New} {Study} of {Bacterial} {Motion}:
  {Superconducting} {Quantum} {Interference} {Device} {Microscopy} of
  {Magnetotactic} {Bacteria}},\ }\href
  {https://doi.org/10.1016/S0006-3495(99)77485-0} {\bibfield  {journal}
  {\bibinfo  {journal} {Biophysical Journal}\ }\textbf {\bibinfo {volume}
  {76}},\ \bibinfo {pages} {3323} (\bibinfo {year} {1999})}\BibitemShut
  {NoStop}%
\bibitem [{\citenamefont {Prozorov}\ \emph {et~al.}(2007)\citenamefont
  {Prozorov}, \citenamefont {Prozorov}, \citenamefont {Mallapragada},
  \citenamefont {Narasimhan}, \citenamefont {Williams},\ and\ \citenamefont
  {Bazylinski}}]{prozorov_magnetic_2007}%
  \BibitemOpen
  \bibfield  {author} {\bibinfo {author} {\bibfnamefont {R.}~\bibnamefont
  {Prozorov}}, \bibinfo {author} {\bibfnamefont {T.}~\bibnamefont {Prozorov}},
  \bibinfo {author} {\bibfnamefont {S.~K.}\ \bibnamefont {Mallapragada}},
  \bibinfo {author} {\bibfnamefont {B.}~\bibnamefont {Narasimhan}}, \bibinfo
  {author} {\bibfnamefont {T.~J.}\ \bibnamefont {Williams}},\ and\ \bibinfo
  {author} {\bibfnamefont {D.~A.}\ \bibnamefont {Bazylinski}},\ }\bibfield
  {title} {\bibinfo {title} {Magnetic irreversibility and the {Verwey}
  transition in nanocrystalline bacterial magnetite},\ }\href
  {https://doi.org/10.1103/PhysRevB.76.054406} {\bibfield  {journal} {\bibinfo
  {journal} {Physical Review B}\ }\textbf {\bibinfo {volume} {76}},\ \bibinfo
  {pages} {054406} (\bibinfo {year} {2007})}\BibitemShut {NoStop}%
\bibitem [{\citenamefont {Lam}\ \emph {et~al.}(2010)\citenamefont {Lam},
  \citenamefont {Hitchcock}, \citenamefont {Obst}, \citenamefont {Lawrence},
  \citenamefont {Swerhone}, \citenamefont {Leppard}, \citenamefont
  {Tyliszczak}, \citenamefont {Karunakaran}, \citenamefont {Wang},
  \citenamefont {Kaznatcheev}, \citenamefont {Bazylinski},\ and\ \citenamefont
  {Lins}}]{lam_characterizing_2010}%
  \BibitemOpen
  \bibfield  {author} {\bibinfo {author} {\bibfnamefont {K.~P.}\ \bibnamefont
  {Lam}}, \bibinfo {author} {\bibfnamefont {A.~P.}\ \bibnamefont {Hitchcock}},
  \bibinfo {author} {\bibfnamefont {M.}~\bibnamefont {Obst}}, \bibinfo {author}
  {\bibfnamefont {J.~R.}\ \bibnamefont {Lawrence}}, \bibinfo {author}
  {\bibfnamefont {G.~D.~W.}\ \bibnamefont {Swerhone}}, \bibinfo {author}
  {\bibfnamefont {G.~G.}\ \bibnamefont {Leppard}}, \bibinfo {author}
  {\bibfnamefont {T.}~\bibnamefont {Tyliszczak}}, \bibinfo {author}
  {\bibfnamefont {C.}~\bibnamefont {Karunakaran}}, \bibinfo {author}
  {\bibfnamefont {J.}~\bibnamefont {Wang}}, \bibinfo {author} {\bibfnamefont
  {K.}~\bibnamefont {Kaznatcheev}}, \bibinfo {author} {\bibfnamefont {D.~A.}\
  \bibnamefont {Bazylinski}},\ and\ \bibinfo {author} {\bibfnamefont
  {U.}~\bibnamefont {Lins}},\ }\bibfield  {title} {\bibinfo {title}
  {Characterizing magnetism of individual magnetosomes by {X}-ray magnetic
  circular dichroism in a scanning transmission {X}-ray microscope},\ }\href
  {https://doi.org/10.1016/j.chemgeo.2009.11.009} {\bibfield  {journal}
  {\bibinfo  {journal} {Chemical Geology}\ }\textbf {\bibinfo {volume} {270}},\
  \bibinfo {pages} {110} (\bibinfo {year} {2010})}\BibitemShut {NoStop}%
\bibitem [{\citenamefont {Marcano}\ \emph {et~al.}(2022)\citenamefont
  {Marcano}, \citenamefont {Orue}, \citenamefont {Gandia}, \citenamefont
  {Gandarias}, \citenamefont {Weigand}, \citenamefont {Abrudan}, \citenamefont
  {Garc\'ia-Prieto}, \citenamefont {Garc\'ia-Arribas}, \citenamefont {Muela},
  \citenamefont {Fern\'andez-Gubieda},\ and\ \citenamefont
  {Valencia}}]{marcano_magnetic_2022}%
  \BibitemOpen
  \bibfield  {author} {\bibinfo {author} {\bibfnamefont {L.}~\bibnamefont
  {Marcano}}, \bibinfo {author} {\bibfnamefont {I.~n.}\ \bibnamefont {Orue}},
  \bibinfo {author} {\bibfnamefont {D.}~\bibnamefont {Gandia}}, \bibinfo
  {author} {\bibfnamefont {L.}~\bibnamefont {Gandarias}}, \bibinfo {author}
  {\bibfnamefont {M.}~\bibnamefont {Weigand}}, \bibinfo {author} {\bibfnamefont
  {R.~M.}\ \bibnamefont {Abrudan}}, \bibinfo {author} {\bibfnamefont
  {A.}~\bibnamefont {Garc\'ia-Prieto}}, \bibinfo {author} {\bibfnamefont
  {A.}~\bibnamefont {Garc\'ia-Arribas}}, \bibinfo {author} {\bibfnamefont
  {A.}~\bibnamefont {Muela}}, \bibinfo {author} {\bibfnamefont {M.~L.}\
  \bibnamefont {Fern\'andez-Gubieda}},\ and\ \bibinfo {author} {\bibfnamefont
  {S.}~\bibnamefont {Valencia}},\ }\bibfield  {title} {\bibinfo {title}
  {Magnetic {Anisotropy} of {Individual} {Nanomagnets} {Embedded} in
  {Biological} {Systems} {Determined} by {Axi}-asymmetric {X}-ray
  {Transmission} {Microscopy}},\ }\href
  {https://doi.org/10.1021/acsnano.1c09559} {\bibfield  {journal} {\bibinfo
  {journal} {ACS Nano}\ }\textbf {\bibinfo {volume} {16}},\ \bibinfo {pages}
  {7398} (\bibinfo {year} {2022})}\BibitemShut {NoStop}%
\bibitem [{\citenamefont {Kalirai}\ \emph {et~al.}(2012)\citenamefont
  {Kalirai}, \citenamefont {Lam}, \citenamefont {Bazylinski}, \citenamefont
  {Lins},\ and\ \citenamefont {Hitchcock}}]{kalirai_examining_2012}%
  \BibitemOpen
  \bibfield  {author} {\bibinfo {author} {\bibfnamefont {S.~S.}\ \bibnamefont
  {Kalirai}}, \bibinfo {author} {\bibfnamefont {K.~P.}\ \bibnamefont {Lam}},
  \bibinfo {author} {\bibfnamefont {D.~A.}\ \bibnamefont {Bazylinski}},
  \bibinfo {author} {\bibfnamefont {U.}~\bibnamefont {Lins}},\ and\ \bibinfo
  {author} {\bibfnamefont {A.~P.}\ \bibnamefont {Hitchcock}},\ }\bibfield
  {title} {\bibinfo {title} {Examining the chemistry and magnetism of
  magnetotactic bacterium {Candidatus} {Magnetovibrio} blakemorei strain {MV}-1
  using scanning transmission {X}-ray microscopy},\ }\href
  {https://doi.org/10.1016/j.chemgeo.2012.01.005} {\bibfield  {journal}
  {\bibinfo  {journal} {Chemical Geology}\ }\textbf {\bibinfo {volume}
  {300-301}},\ \bibinfo {pages} {14} (\bibinfo {year} {2012})}\BibitemShut
  {NoStop}%
\bibitem [{\citenamefont {Kalirai}\ \emph {et~al.}(2013)\citenamefont
  {Kalirai}, \citenamefont {Bazylinski},\ and\ \citenamefont
  {Hitchcock}}]{kalirai_anomalous_2013}%
  \BibitemOpen
  \bibfield  {author} {\bibinfo {author} {\bibfnamefont {S.~S.}\ \bibnamefont
  {Kalirai}}, \bibinfo {author} {\bibfnamefont {D.~A.}\ \bibnamefont
  {Bazylinski}},\ and\ \bibinfo {author} {\bibfnamefont {A.~P.}\ \bibnamefont
  {Hitchcock}},\ }\bibfield  {title} {\bibinfo {title} {Anomalous {Magnetic}
  {Orientations} of {Magnetosome} {Chains} in a {Magnetotactic} {Bacterium}:
  {Magnetovibrio} blakemorei {Strain} {MV}-1},\ }\href
  {https://doi.org/10.1371/journal.pone.0053368} {\bibfield  {journal}
  {\bibinfo  {journal} {PLOS ONE}\ }\textbf {\bibinfo {volume} {8}},\ \bibinfo
  {pages} {e53368} (\bibinfo {year} {2013})}\BibitemShut {NoStop}%
\bibitem [{\citenamefont {Staniland}\ \emph {et~al.}(2007)\citenamefont
  {Staniland}, \citenamefont {Ward}, \citenamefont {Harrison}, \citenamefont
  {van~der Laan},\ and\ \citenamefont {Telling}}]{staniland_rapid_2007}%
  \BibitemOpen
  \bibfield  {author} {\bibinfo {author} {\bibfnamefont {S.}~\bibnamefont
  {Staniland}}, \bibinfo {author} {\bibfnamefont {B.}~\bibnamefont {Ward}},
  \bibinfo {author} {\bibfnamefont {A.}~\bibnamefont {Harrison}}, \bibinfo
  {author} {\bibfnamefont {G.}~\bibnamefont {van~der Laan}},\ and\ \bibinfo
  {author} {\bibfnamefont {N.}~\bibnamefont {Telling}},\ }\bibfield  {title}
  {\bibinfo {title} {Rapid magnetosome formation shown by real-time x-ray
  magnetic circular dichroism},\ }\href
  {https://doi.org/10.1073/pnas.0704879104} {\bibfield  {journal} {\bibinfo
  {journal} {Proceedings of the National Academy of Sciences}\ }\textbf
  {\bibinfo {volume} {104}},\ \bibinfo {pages} {19524} (\bibinfo {year}
  {2007})}\BibitemShut {NoStop}%
\bibitem [{\citenamefont {Dunin-Borkowski}\ \emph {et~al.}(1998)\citenamefont
  {Dunin-Borkowski}, \citenamefont {McCartney}, \citenamefont {Frankel},
  \citenamefont {Bazylinski}, \citenamefont {Pósfai},\ and\ \citenamefont
  {Buseck}}]{dunin-borkowski_magnetic_1998}%
  \BibitemOpen
  \bibfield  {author} {\bibinfo {author} {\bibfnamefont {R.~E.}\ \bibnamefont
  {Dunin-Borkowski}}, \bibinfo {author} {\bibfnamefont {M.~R.}\ \bibnamefont
  {McCartney}}, \bibinfo {author} {\bibfnamefont {R.~B.}\ \bibnamefont
  {Frankel}}, \bibinfo {author} {\bibfnamefont {D.~A.}\ \bibnamefont
  {Bazylinski}}, \bibinfo {author} {\bibfnamefont {M.}~\bibnamefont
  {Pósfai}},\ and\ \bibinfo {author} {\bibfnamefont {P.~R.}\ \bibnamefont
  {Buseck}},\ }\bibfield  {title} {\bibinfo {title} {Magnetic {Microstructure}
  of {Magnetotactic} {Bacteria} by {Electron} {Holography}},\ }\href
  {https://doi.org/10.1126/science.282.5395.1868} {\bibfield  {journal}
  {\bibinfo  {journal} {Science}\ }\textbf {\bibinfo {volume} {282}},\ \bibinfo
  {pages} {1868} (\bibinfo {year} {1998})}\BibitemShut {NoStop}%
\bibitem [{\citenamefont {Wasem}(2015)}]{wasem_atomic_2015}%
  \BibitemOpen
  \bibfield  {author} {\bibinfo {author} {\bibfnamefont {M.}~\bibnamefont
  {Wasem}},\ }\emph {\bibinfo {title} {Atomic {Force} {Microscopy} of
  {Nanoparticles} and {Biological} {Cells}}},\ \href
  {https://doi.org/10.5451/unibas-006492030} {\bibinfo {type} {Thesis}},\
  \bibinfo  {school} {University of Basel} (\bibinfo {year} {2015})\BibitemShut
  {NoStop}%
\bibitem [{\citenamefont {Orue}\ \emph {et~al.}(2018)\citenamefont {Orue},
  \citenamefont {Marcano}, \citenamefont {Bender}, \citenamefont
  {Garc\'ia-Prieto}, \citenamefont {Valencia}, \citenamefont {Mawass},
  \citenamefont {Gil-Cart\'on}, \citenamefont {Venero}, \citenamefont
  {Honecker}, \citenamefont {Garc\'ia-Arribas}, \citenamefont {Barqu\'in},
  \citenamefont {Muela},\ and\ \citenamefont
  {Fern\'andez-Gubieda}}]{orue_configuration_2018}%
  \BibitemOpen
  \bibfield  {author} {\bibinfo {author} {\bibfnamefont {I.}~\bibnamefont
  {Orue}}, \bibinfo {author} {\bibfnamefont {L.}~\bibnamefont {Marcano}},
  \bibinfo {author} {\bibfnamefont {P.}~\bibnamefont {Bender}}, \bibinfo
  {author} {\bibfnamefont {A.}~\bibnamefont {Garc\'ia-Prieto}}, \bibinfo
  {author} {\bibfnamefont {S.}~\bibnamefont {Valencia}}, \bibinfo {author}
  {\bibfnamefont {M.~A.}\ \bibnamefont {Mawass}}, \bibinfo {author}
  {\bibfnamefont {D.}~\bibnamefont {Gil-Cart\'on}}, \bibinfo {author}
  {\bibfnamefont {D.~A.}\ \bibnamefont {Venero}}, \bibinfo {author}
  {\bibfnamefont {D.}~\bibnamefont {Honecker}}, \bibinfo {author}
  {\bibfnamefont {A.}~\bibnamefont {Garc\'ia-Arribas}}, \bibinfo {author}
  {\bibfnamefont {L.~F.}\ \bibnamefont {Barqu\'in}}, \bibinfo {author}
  {\bibfnamefont {A.}~\bibnamefont {Muela}},\ and\ \bibinfo {author}
  {\bibfnamefont {M.~L.}\ \bibnamefont {Fern\'andez-Gubieda}},\ }\bibfield
  {title} {\bibinfo {title} {Configuration of the magnetosome chain: a natural
  magnetic nanoarchitecture},\ }\href {https://doi.org/10.1039/C7NR08493E}
  {\bibfield  {journal} {\bibinfo  {journal} {Nanoscale}\ }\textbf {\bibinfo
  {volume} {10}},\ \bibinfo {pages} {7407} (\bibinfo {year}
  {2018})}\BibitemShut {NoStop}%
\bibitem [{\citenamefont {Rossel}\ \emph {et~al.}(1996)\citenamefont {Rossel},
  \citenamefont {Bauer}, \citenamefont {Zech}, \citenamefont {Hofer},
  \citenamefont {Willemin},\ and\ \citenamefont {Keller}}]{rossel_active_1996}%
  \BibitemOpen
  \bibfield  {author} {\bibinfo {author} {\bibfnamefont {C.}~\bibnamefont
  {Rossel}}, \bibinfo {author} {\bibfnamefont {P.}~\bibnamefont {Bauer}},
  \bibinfo {author} {\bibfnamefont {D.}~\bibnamefont {Zech}}, \bibinfo {author}
  {\bibfnamefont {J.}~\bibnamefont {Hofer}}, \bibinfo {author} {\bibfnamefont
  {M.}~\bibnamefont {Willemin}},\ and\ \bibinfo {author} {\bibfnamefont
  {H.}~\bibnamefont {Keller}},\ }\bibfield  {title} {\bibinfo {title} {Active
  microlevers as miniature torque magnetometers},\ }\href
  {https://doi.org/10.1063/1.362550} {\bibfield  {journal} {\bibinfo  {journal}
  {Journal of Applied Physics}\ }\textbf {\bibinfo {volume} {79}},\ \bibinfo
  {pages} {8166} (\bibinfo {year} {1996})}\BibitemShut {NoStop}%
\bibitem [{\citenamefont {Harris}\ \emph {et~al.}(1999)\citenamefont {Harris},
  \citenamefont {Awschalom}, \citenamefont {Matsukura}, \citenamefont {Ohno},
  \citenamefont {Maranowski},\ and\ \citenamefont
  {Gossard}}]{harris_integrated_1999}%
  \BibitemOpen
  \bibfield  {author} {\bibinfo {author} {\bibfnamefont {J.~G.~E.}\
  \bibnamefont {Harris}}, \bibinfo {author} {\bibfnamefont {D.~D.}\
  \bibnamefont {Awschalom}}, \bibinfo {author} {\bibfnamefont {F.}~\bibnamefont
  {Matsukura}}, \bibinfo {author} {\bibfnamefont {H.}~\bibnamefont {Ohno}},
  \bibinfo {author} {\bibfnamefont {K.~D.}\ \bibnamefont {Maranowski}},\ and\
  \bibinfo {author} {\bibfnamefont {A.~C.}\ \bibnamefont {Gossard}},\
  }\bibfield  {title} {\bibinfo {title} {Integrated micromechanical cantilever
  magnetometry of {Ga1}-{xMnxAs}},\ }\href {https://doi.org/10.1063/1.124622}
  {\bibfield  {journal} {\bibinfo  {journal} {Applied Physics Letters}\
  }\textbf {\bibinfo {volume} {75}},\ \bibinfo {pages} {1140} (\bibinfo {year}
  {1999})}\BibitemShut {NoStop}%
\bibitem [{\citenamefont {Stipe}\ \emph {et~al.}(2001)\citenamefont {Stipe},
  \citenamefont {Mamin}, \citenamefont {Stowe}, \citenamefont {Kenny},\ and\
  \citenamefont {Rugar}}]{stipe_magnetic_2001}%
  \BibitemOpen
  \bibfield  {author} {\bibinfo {author} {\bibfnamefont {B.~C.}\ \bibnamefont
  {Stipe}}, \bibinfo {author} {\bibfnamefont {H.~J.}\ \bibnamefont {Mamin}},
  \bibinfo {author} {\bibfnamefont {T.~D.}\ \bibnamefont {Stowe}}, \bibinfo
  {author} {\bibfnamefont {T.~W.}\ \bibnamefont {Kenny}},\ and\ \bibinfo
  {author} {\bibfnamefont {D.}~\bibnamefont {Rugar}},\ }\bibfield  {title}
  {\bibinfo {title} {Magnetic {Dissipation} and {Fluctuations} in {Individual}
  {Nanomagnets} {Measured} by {Ultrasensitive} {Cantilever} {Magnetometry}},\
  }\href {https://doi.org/10.1103/PhysRevLett.86.2874} {\bibfield  {journal}
  {\bibinfo  {journal} {Physical Review Letters}\ }\textbf {\bibinfo {volume}
  {86}},\ \bibinfo {pages} {2874} (\bibinfo {year} {2001})}\BibitemShut
  {NoStop}%
\bibitem [{\citenamefont {Gross}\ \emph {et~al.}(2016)\citenamefont {Gross},
  \citenamefont {Weber}, \citenamefont {Rüffer}, \citenamefont {Buchter},
  \citenamefont {Heimbach}, \citenamefont {Fontcuberta~i Morral}, \citenamefont
  {Grundler},\ and\ \citenamefont {Poggio}}]{gross_dynamic_2016}%
  \BibitemOpen
  \bibfield  {author} {\bibinfo {author} {\bibfnamefont {B.}~\bibnamefont
  {Gross}}, \bibinfo {author} {\bibfnamefont {D.~P.}\ \bibnamefont {Weber}},
  \bibinfo {author} {\bibfnamefont {D.}~\bibnamefont {Rüffer}}, \bibinfo
  {author} {\bibfnamefont {A.}~\bibnamefont {Buchter}}, \bibinfo {author}
  {\bibfnamefont {F.}~\bibnamefont {Heimbach}}, \bibinfo {author}
  {\bibfnamefont {A.}~\bibnamefont {Fontcuberta~i Morral}}, \bibinfo {author}
  {\bibfnamefont {D.}~\bibnamefont {Grundler}},\ and\ \bibinfo {author}
  {\bibfnamefont {M.}~\bibnamefont {Poggio}},\ }\bibfield  {title} {\bibinfo
  {title} {Dynamic cantilever magnetometry of individual {CoFeB} nanotubes},\
  }\href {https://doi.org/10.1103/PhysRevB.93.064409} {\bibfield  {journal}
  {\bibinfo  {journal} {Physical Review B}\ }\textbf {\bibinfo {volume} {93}},\
  \bibinfo {pages} {064409} (\bibinfo {year} {2016})}\BibitemShut {NoStop}%
\bibitem [{\citenamefont {Gysin}\ \emph {et~al.}(2011)\citenamefont {Gysin},
  \citenamefont {Rast}, \citenamefont {Aste}, \citenamefont {Speliotis},
  \citenamefont {Werle},\ and\ \citenamefont {Meyer}}]{gysin_magnetic_2011}%
  \BibitemOpen
  \bibfield  {author} {\bibinfo {author} {\bibfnamefont {U.}~\bibnamefont
  {Gysin}}, \bibinfo {author} {\bibfnamefont {S.}~\bibnamefont {Rast}},
  \bibinfo {author} {\bibfnamefont {A.}~\bibnamefont {Aste}}, \bibinfo {author}
  {\bibfnamefont {T.}~\bibnamefont {Speliotis}}, \bibinfo {author}
  {\bibfnamefont {C.}~\bibnamefont {Werle}},\ and\ \bibinfo {author}
  {\bibfnamefont {E.}~\bibnamefont {Meyer}},\ }\bibfield  {title} {\bibinfo
  {title} {Magnetic properties of nanomagnetic and biomagnetic systems analyzed
  using cantilever magnetometry},\ }\href
  {https://doi.org/10.1088/0957-4484/22/28/285715} {\bibfield  {journal}
  {\bibinfo  {journal} {Nanotechnology}\ }\textbf {\bibinfo {volume} {22}},\
  \bibinfo {pages} {285715} (\bibinfo {year} {2011})}\BibitemShut {NoStop}%
\bibitem [{\citenamefont {Modic}\ \emph {et~al.}(2018)\citenamefont {Modic},
  \citenamefont {Bachmann}, \citenamefont {Ramshaw}, \citenamefont {Arnold},
  \citenamefont {Shirer}, \citenamefont {Estry}, \citenamefont {Betts},
  \citenamefont {Ghimire}, \citenamefont {Bauer}, \citenamefont {Schmidt},
  \citenamefont {Baenitz}, \citenamefont {Svanidze}, \citenamefont {McDonald},
  \citenamefont {Shekhter},\ and\ \citenamefont {Moll}}]{modic_resonant_2018}%
  \BibitemOpen
  \bibfield  {author} {\bibinfo {author} {\bibfnamefont {K.~A.}\ \bibnamefont
  {Modic}}, \bibinfo {author} {\bibfnamefont {M.~D.}\ \bibnamefont {Bachmann}},
  \bibinfo {author} {\bibfnamefont {B.~J.}\ \bibnamefont {Ramshaw}}, \bibinfo
  {author} {\bibfnamefont {F.}~\bibnamefont {Arnold}}, \bibinfo {author}
  {\bibfnamefont {K.~R.}\ \bibnamefont {Shirer}}, \bibinfo {author}
  {\bibfnamefont {A.}~\bibnamefont {Estry}}, \bibinfo {author} {\bibfnamefont
  {J.~B.}\ \bibnamefont {Betts}}, \bibinfo {author} {\bibfnamefont {N.~J.}\
  \bibnamefont {Ghimire}}, \bibinfo {author} {\bibfnamefont {E.~D.}\
  \bibnamefont {Bauer}}, \bibinfo {author} {\bibfnamefont {M.}~\bibnamefont
  {Schmidt}}, \bibinfo {author} {\bibfnamefont {M.}~\bibnamefont {Baenitz}},
  \bibinfo {author} {\bibfnamefont {E.}~\bibnamefont {Svanidze}}, \bibinfo
  {author} {\bibfnamefont {R.~D.}\ \bibnamefont {McDonald}}, \bibinfo {author}
  {\bibfnamefont {A.}~\bibnamefont {Shekhter}},\ and\ \bibinfo {author}
  {\bibfnamefont {P.~J.~W.}\ \bibnamefont {Moll}},\ }\bibfield  {title}
  {\bibinfo {title} {Resonant torsion magnetometry in anisotropic quantum
  materials},\ }\href {https://doi.org/10.1038/s41467-018-06412-w} {\bibfield
  {journal} {\bibinfo  {journal} {Nature Communications}\ }\textbf {\bibinfo
  {volume} {9}},\ \bibinfo {pages} {3975} (\bibinfo {year} {2018})}\BibitemShut
  {NoStop}%
\bibitem [{\citenamefont {Gross}\ \emph {et~al.}(2021)\citenamefont {Gross},
  \citenamefont {Philipp}, \citenamefont {Josten}, \citenamefont {Leliaert},
  \citenamefont {Wetterskog}, \citenamefont {Bergström},\ and\ \citenamefont
  {Poggio}}]{gross_magnetic_2021}%
  \BibitemOpen
  \bibfield  {author} {\bibinfo {author} {\bibfnamefont {B.}~\bibnamefont
  {Gross}}, \bibinfo {author} {\bibfnamefont {S.}~\bibnamefont {Philipp}},
  \bibinfo {author} {\bibfnamefont {E.}~\bibnamefont {Josten}}, \bibinfo
  {author} {\bibfnamefont {J.}~\bibnamefont {Leliaert}}, \bibinfo {author}
  {\bibfnamefont {E.}~\bibnamefont {Wetterskog}}, \bibinfo {author}
  {\bibfnamefont {L.}~\bibnamefont {Bergström}},\ and\ \bibinfo {author}
  {\bibfnamefont {M.}~\bibnamefont {Poggio}},\ }\bibfield  {title} {\bibinfo
  {title} {Magnetic anisotropy of individual maghemite mesocrystals},\ }\href
  {https://doi.org/10.1103/PhysRevB.103.014402} {\bibfield  {journal} {\bibinfo
   {journal} {Physical Review B}\ }\textbf {\bibinfo {volume} {103}},\ \bibinfo
  {pages} {014402} (\bibinfo {year} {2021})}\BibitemShut {NoStop}%
\bibitem [{\citenamefont {Mehlin}\ \emph {et~al.}(2015)\citenamefont {Mehlin},
  \citenamefont {Xue}, \citenamefont {Liang}, \citenamefont {Du}, \citenamefont
  {Stolt}, \citenamefont {Jin}, \citenamefont {Tian},\ and\ \citenamefont
  {Poggio}}]{mehlin_stabilized_2015}%
  \BibitemOpen
  \bibfield  {author} {\bibinfo {author} {\bibfnamefont {A.}~\bibnamefont
  {Mehlin}}, \bibinfo {author} {\bibfnamefont {F.}~\bibnamefont {Xue}},
  \bibinfo {author} {\bibfnamefont {D.}~\bibnamefont {Liang}}, \bibinfo
  {author} {\bibfnamefont {H.~F.}\ \bibnamefont {Du}}, \bibinfo {author}
  {\bibfnamefont {M.~J.}\ \bibnamefont {Stolt}}, \bibinfo {author}
  {\bibfnamefont {S.}~\bibnamefont {Jin}}, \bibinfo {author} {\bibfnamefont
  {M.~L.}\ \bibnamefont {Tian}},\ and\ \bibinfo {author} {\bibfnamefont
  {M.}~\bibnamefont {Poggio}},\ }\bibfield  {title} {\bibinfo {title}
  {Stabilized {Skyrmion} {Phase} {Detected} in {MnSi} {Nanowires} by {Dynamic}
  {Cantilever} {Magnetometry}},\ }\href
  {https://doi.org/10.1021/acs.nanolett.5b02232} {\bibfield  {journal}
  {\bibinfo  {journal} {Nano Letters}\ }\textbf {\bibinfo {volume} {15}},\
  \bibinfo {pages} {4839} (\bibinfo {year} {2015})}\BibitemShut {NoStop}%
\bibitem [{\citenamefont {Gross}\ \emph {et~al.}(2020)\citenamefont {Gross},
  \citenamefont {Philipp}, \citenamefont {Geirhos}, \citenamefont {Mehlin},
  \citenamefont {Bordács}, \citenamefont {Tsurkan}, \citenamefont {Leonov},
  \citenamefont {Kézsmárki},\ and\ \citenamefont
  {Poggio}}]{gross_stability_2020}%
  \BibitemOpen
  \bibfield  {author} {\bibinfo {author} {\bibfnamefont {B.}~\bibnamefont
  {Gross}}, \bibinfo {author} {\bibfnamefont {S.}~\bibnamefont {Philipp}},
  \bibinfo {author} {\bibfnamefont {K.}~\bibnamefont {Geirhos}}, \bibinfo
  {author} {\bibfnamefont {A.}~\bibnamefont {Mehlin}}, \bibinfo {author}
  {\bibfnamefont {S.}~\bibnamefont {Bordács}}, \bibinfo {author}
  {\bibfnamefont {V.}~\bibnamefont {Tsurkan}}, \bibinfo {author} {\bibfnamefont
  {A.}~\bibnamefont {Leonov}}, \bibinfo {author} {\bibfnamefont
  {I.}~\bibnamefont {Kézsmárki}},\ and\ \bibinfo {author} {\bibfnamefont
  {M.}~\bibnamefont {Poggio}},\ }\bibfield  {title} {\bibinfo {title}
  {Stability of {N}\'eel-type skyrmion lattice against oblique magnetic fields
  in {Ga}{V}$_4${S}$_8$ and {Ga}{V}$_4${Se}$_8$},\ }\href
  {https://doi.org/10.1103/PhysRevB.102.104407} {\bibfield  {journal} {\bibinfo
   {journal} {Physical Review B}\ }\textbf {\bibinfo {volume} {102}},\ \bibinfo
  {pages} {104407} (\bibinfo {year} {2020})}\BibitemShut {NoStop}%
\bibitem [{\citenamefont {Philipp}\ \emph {et~al.}(2021)\citenamefont
  {Philipp}, \citenamefont {Gross}, \citenamefont {Reginka}, \citenamefont
  {Merkel}, \citenamefont {Claus}, \citenamefont {Sulliger}, \citenamefont
  {Ehresmann},\ and\ \citenamefont {Poggio}}]{philipp_magnetic_2021}%
  \BibitemOpen
  \bibfield  {author} {\bibinfo {author} {\bibfnamefont {S.}~\bibnamefont
  {Philipp}}, \bibinfo {author} {\bibfnamefont {B.}~\bibnamefont {Gross}},
  \bibinfo {author} {\bibfnamefont {M.}~\bibnamefont {Reginka}}, \bibinfo
  {author} {\bibfnamefont {M.}~\bibnamefont {Merkel}}, \bibinfo {author}
  {\bibfnamefont {M.~M.}\ \bibnamefont {Claus}}, \bibinfo {author}
  {\bibfnamefont {M.}~\bibnamefont {Sulliger}}, \bibinfo {author}
  {\bibfnamefont {A.}~\bibnamefont {Ehresmann}},\ and\ \bibinfo {author}
  {\bibfnamefont {M.}~\bibnamefont {Poggio}},\ }\bibfield  {title} {\bibinfo
  {title} {Magnetic hysteresis of individual {Janus} particles with
  hemispherical exchange biased caps},\ }\href
  {https://doi.org/10.1063/5.0076116} {\bibfield  {journal} {\bibinfo
  {journal} {Applied Physics Letters}\ }\textbf {\bibinfo {volume} {119}},\
  \bibinfo {pages} {222406} (\bibinfo {year} {2021})}\BibitemShut {NoStop}%
\bibitem [{\citenamefont {Rugar}(1989)}]{rugar_improved_1989}%
  \BibitemOpen
  \bibfield  {author} {\bibinfo {author} {\bibfnamefont {D.}~\bibnamefont
  {Rugar}},\ }\bibfield  {title} {\bibinfo {title} {Improved fiber‐optic
  interferometer for atomic force microscopy},\ }\href
  {https://doi.org/10.1063/1.101987} {\bibfield  {journal} {\bibinfo  {journal}
  {Applied Physics Letters}\ }\textbf {\bibinfo {volume} {55}},\ \bibinfo
  {pages} {2588} (\bibinfo {year} {1989})}\BibitemShut {NoStop}%
\bibitem [{SOM()}]{SOM}%
  \BibitemOpen
  \href@noop {} {\ }\bibinfo {note} {See Supplemental Material [url] for
  further information.}\BibitemShut {Stop}%
\bibitem [{\citenamefont {Heyen}\ and\ \citenamefont
  {Schüler}(2003)}]{heyen_growth_2003}%
  \BibitemOpen
  \bibfield  {author} {\bibinfo {author} {\bibfnamefont {U.}~\bibnamefont
  {Heyen}}\ and\ \bibinfo {author} {\bibfnamefont {D.}~\bibnamefont
  {Schüler}},\ }\bibfield  {title} {\bibinfo {title} {Growth and magnetosome
  formation by microaerophilic {Magnetospirillum} strains in an
  oxygen-controlled fermentor},\ }\href
  {https://doi.org/10.1007/s00253-002-1219-x} {\bibfield  {journal} {\bibinfo
  {journal} {Applied Microbiology and Biotechnology}\ }\textbf {\bibinfo
  {volume} {61}},\ \bibinfo {pages} {536} (\bibinfo {year} {2003})}\BibitemShut
  {NoStop}%
\bibitem [{\citenamefont {Li}\ \emph {et~al.}(2012)\citenamefont {Li},
  \citenamefont {Katzmann}, \citenamefont {Borg},\ and\ \citenamefont
  {Schüler}}]{li_periplasmic_2012}%
  \BibitemOpen
  \bibfield  {author} {\bibinfo {author} {\bibfnamefont {Y.}~\bibnamefont
  {Li}}, \bibinfo {author} {\bibfnamefont {E.}~\bibnamefont {Katzmann}},
  \bibinfo {author} {\bibfnamefont {S.}~\bibnamefont {Borg}},\ and\ \bibinfo
  {author} {\bibfnamefont {D.}~\bibnamefont {Schüler}},\ }\bibfield  {title}
  {\bibinfo {title} {The {Periplasmic} {Nitrate} {Reductase} {Nap} {Is}
  {Required} for {Anaerobic} {Growth} and {Involved} in {Redox} {Control} of
  {Magnetite} {Biomineralization} in {Magnetospirillum} gryphiswaldense},\
  }\href {https://doi.org/10.1128/jb.00903-12} {\bibfield  {journal} {\bibinfo
  {journal} {Journal of Bacteriology}\ }\textbf {\bibinfo {volume} {194}},\
  \bibinfo {pages} {4847} (\bibinfo {year} {2012})}\BibitemShut {NoStop}%
\bibitem [{\citenamefont {Vansteenkiste}\ \emph {et~al.}(2014)\citenamefont
  {Vansteenkiste}, \citenamefont {Leliaert}, \citenamefont {Dvornik},
  \citenamefont {Helsen}, \citenamefont {Garcia-Sanchez},\ and\ \citenamefont
  {Van~Waeyenberge}}]{vansteenkiste_design_2014}%
  \BibitemOpen
  \bibfield  {author} {\bibinfo {author} {\bibfnamefont {A.}~\bibnamefont
  {Vansteenkiste}}, \bibinfo {author} {\bibfnamefont {J.}~\bibnamefont
  {Leliaert}}, \bibinfo {author} {\bibfnamefont {M.}~\bibnamefont {Dvornik}},
  \bibinfo {author} {\bibfnamefont {M.}~\bibnamefont {Helsen}}, \bibinfo
  {author} {\bibfnamefont {F.}~\bibnamefont {Garcia-Sanchez}},\ and\ \bibinfo
  {author} {\bibfnamefont {B.}~\bibnamefont {Van~Waeyenberge}},\ }\bibfield
  {title} {\bibinfo {title} {The design and verification of {MuMax3}},\ }\href
  {https://doi.org/10.1063/1.4899186} {\bibfield  {journal} {\bibinfo
  {journal} {AIP Advances}\ }\textbf {\bibinfo {volume} {4}},\ \bibinfo {pages}
  {107133} (\bibinfo {year} {2014})}\BibitemShut {NoStop}%
\bibitem [{\citenamefont {Exl}\ \emph {et~al.}(2014)\citenamefont {Exl},
  \citenamefont {Bance}, \citenamefont {Reichel}, \citenamefont {Schrefl},
  \citenamefont {Peter~Stimming},\ and\ \citenamefont
  {Mauser}}]{exl_labontes_2014}%
  \BibitemOpen
  \bibfield  {author} {\bibinfo {author} {\bibfnamefont {L.}~\bibnamefont
  {Exl}}, \bibinfo {author} {\bibfnamefont {S.}~\bibnamefont {Bance}}, \bibinfo
  {author} {\bibfnamefont {F.}~\bibnamefont {Reichel}}, \bibinfo {author}
  {\bibfnamefont {T.}~\bibnamefont {Schrefl}}, \bibinfo {author} {\bibfnamefont
  {H.}~\bibnamefont {Peter~Stimming}},\ and\ \bibinfo {author} {\bibfnamefont
  {N.~J.}\ \bibnamefont {Mauser}},\ }\bibfield  {title} {\bibinfo {title}
  {{LaBonte}'s method revisited: {An} effective steepest descent method for
  micromagnetic energy minimization},\ }\href
  {https://doi.org/10.1063/1.4862839} {\bibfield  {journal} {\bibinfo
  {journal} {Journal of Applied Physics}\ }\textbf {\bibinfo {volume} {115}},\
  \bibinfo {pages} {17D118} (\bibinfo {year} {2014})}\BibitemShut {NoStop}%
\bibitem [{\citenamefont {Ó~Conbhuí}\ \emph {et~al.}(2018)\citenamefont
  {Ó~Conbhuí}, \citenamefont {Williams}, \citenamefont {Fabian},
  \citenamefont {Ridley}, \citenamefont {Nagy},\ and\ \citenamefont
  {Muxworthy}}]{o_conbhui_merrill_2018}%
  \BibitemOpen
  \bibfield  {author} {\bibinfo {author} {\bibfnamefont {P.}~\bibnamefont
  {Ó~Conbhuí}}, \bibinfo {author} {\bibfnamefont {W.}~\bibnamefont
  {Williams}}, \bibinfo {author} {\bibfnamefont {K.}~\bibnamefont {Fabian}},
  \bibinfo {author} {\bibfnamefont {P.}~\bibnamefont {Ridley}}, \bibinfo
  {author} {\bibfnamefont {L.}~\bibnamefont {Nagy}},\ and\ \bibinfo {author}
  {\bibfnamefont {A.~R.}\ \bibnamefont {Muxworthy}},\ }\bibfield  {title}
  {\bibinfo {title} {{MERRILL}: {Micromagnetic} {Earth} {Related} {Robust}
  {Interpreted} {Language} {Laboratory}},\ }\href
  {https://doi.org/10.1002/2017GC007279} {\bibfield  {journal} {\bibinfo
  {journal} {Geochemistry, Geophysics, Geosystems}\ }\textbf {\bibinfo {volume}
  {19}},\ \bibinfo {pages} {1080} (\bibinfo {year} {2018})}\BibitemShut
  {NoStop}%
\bibitem [{\citenamefont {Gandia}\ \emph {et~al.}(2020)\citenamefont {Gandia},
  \citenamefont {Gandarias}, \citenamefont {Marcano}, \citenamefont {Orue},
  \citenamefont {Gil-Cart\'on}, \citenamefont {Alonso}, \citenamefont
  {Garc\'ia-Arribas}, \citenamefont {Muela},\ and\ \citenamefont
  {Fern\'andez-Gubieda}}]{gandia_elucidating_2020}%
  \BibitemOpen
  \bibfield  {author} {\bibinfo {author} {\bibfnamefont {D.}~\bibnamefont
  {Gandia}}, \bibinfo {author} {\bibfnamefont {L.}~\bibnamefont {Gandarias}},
  \bibinfo {author} {\bibfnamefont {L.}~\bibnamefont {Marcano}}, \bibinfo
  {author} {\bibfnamefont {I.~n.}\ \bibnamefont {Orue}}, \bibinfo {author}
  {\bibfnamefont {D.}~\bibnamefont {Gil-Cart\'on}}, \bibinfo {author}
  {\bibfnamefont {J.}~\bibnamefont {Alonso}}, \bibinfo {author} {\bibfnamefont
  {A.}~\bibnamefont {Garc\'ia-Arribas}}, \bibinfo {author} {\bibfnamefont
  {A.}~\bibnamefont {Muela}},\ and\ \bibinfo {author} {\bibfnamefont {M.~L.}\
  \bibnamefont {Fern\'andez-Gubieda}},\ }\bibfield  {title} {\bibinfo {title}
  {Elucidating the role of shape anisotropy in faceted magnetic nanoparticles
  using biogenic magnetosomes as a model},\ }\href
  {https://doi.org/10.1039/D0NR02189J} {\bibfield  {journal} {\bibinfo
  {journal} {Nanoscale}\ }\textbf {\bibinfo {volume} {12}},\ \bibinfo {pages}
  {16081} (\bibinfo {year} {2020})}\BibitemShut {NoStop}%
\bibitem [{\citenamefont {Körnig}\ \emph {et~al.}(2014)\citenamefont
  {Körnig}, \citenamefont {Winklhofer}, \citenamefont {Baumgartner},
  \citenamefont {Gonzalez}, \citenamefont {Fratzl},\ and\ \citenamefont
  {Faivre}}]{kornig_magnetite_2014}%
  \BibitemOpen
  \bibfield  {author} {\bibinfo {author} {\bibfnamefont {A.}~\bibnamefont
  {Körnig}}, \bibinfo {author} {\bibfnamefont {M.}~\bibnamefont {Winklhofer}},
  \bibinfo {author} {\bibfnamefont {J.}~\bibnamefont {Baumgartner}}, \bibinfo
  {author} {\bibfnamefont {T.~P.}\ \bibnamefont {Gonzalez}}, \bibinfo {author}
  {\bibfnamefont {P.}~\bibnamefont {Fratzl}},\ and\ \bibinfo {author}
  {\bibfnamefont {D.}~\bibnamefont {Faivre}},\ }\bibfield  {title} {\bibinfo
  {title} {Magnetite {Crystal} {Orientation} in {Magnetosome} {Chains}},\
  }\href {https://doi.org/10.1002/adfm.201303737} {\bibfield  {journal}
  {\bibinfo  {journal} {Advanced Functional Materials}\ }\textbf {\bibinfo
  {volume} {24}},\ \bibinfo {pages} {3926} (\bibinfo {year}
  {2014})}\BibitemShut {NoStop}%
\bibitem [{\citenamefont {Mann}\ \emph {et~al.}(1984)\citenamefont {Mann},
  \citenamefont {Frankel},\ and\ \citenamefont
  {Blakemore}}]{mann_structure_1984}%
  \BibitemOpen
  \bibfield  {author} {\bibinfo {author} {\bibfnamefont {S.}~\bibnamefont
  {Mann}}, \bibinfo {author} {\bibfnamefont {R.~B.}\ \bibnamefont {Frankel}},\
  and\ \bibinfo {author} {\bibfnamefont {R.~P.}\ \bibnamefont {Blakemore}},\
  }\bibfield  {title} {\bibinfo {title} {Structure, morphology and crystal
  growth of bacterial magnetite},\ }\href {https://doi.org/10.1038/310405a0}
  {\bibfield  {journal} {\bibinfo  {journal} {Nature}\ }\textbf {\bibinfo
  {volume} {310}},\ \bibinfo {pages} {405} (\bibinfo {year}
  {1984})}\BibitemShut {NoStop}%
\end{thebibliography}
%

%
\end{document}